\documentclass[journal]{IEEEtran}
\usepackage{cite}
\usepackage{amsmath,amssymb,amsfonts}
\usepackage{algorithmic}
\usepackage{graphicx}
\usepackage{textcomp}
\usepackage{xcolor}
\usepackage[lofdepth,lotdepth]{subfig}
\usepackage{caption}
\usepackage[inline]{enumitem}
\usepackage{amsmath}
\usepackage{alphabeta}
\usepackage{threeparttable}
\usepackage{hyperref}

\usepackage[nolist]{acronym} % SB
\usepackage{multirow}
\usepackage[draft]{changes}  % https://ctan.org/pkg/changes?lang=en
\ifCLASSINFOpdf

\else

\fi

\usepackage{lipsum}
\usepackage{fancyhdr}

\fancypagestyle{sec}{\lhead{\tmpx}}

\fancypagestyle{arXiv}
{
   \fancyhf{}
   \chead{\textcolor{gray}{(Invited Paper) This paper has been accepted for publication at the \\
   IEEE Journal on Emerging and Selected Topics in Circuits and Systems, vol. 10, no. 4, pp. 419-432, Dec. 2020}}
   \cfoot{\thepage}
   
}

\begin{document}
\title{EdgeDRNN: Recurrent Neural Network Accelerator for Edge Inference}

% \author{Michael~Shell,~\IEEEmembership{Member,~IEEE,}
%         John~Doe,~\IEEEmembership{Fellow,~OSA,}
%         and~Jane~Doe,~\IEEEmembership{Life~Fellow,~IEEE}% <-this % stops a space

\author{
    Chang Gao,~\IEEEmembership{Student~Member,~IEEE,}
    Antonio Rios-Navarro,~\IEEEmembership{Member,~IEEE,}
    Xi Chen,~\IEEEmembership{Student~Member,~IEEE}
    Shih-Chii Liu,~\IEEEmembership{Senior~Member,~IEEE}
    Tobi Delbruck,~\IEEEmembership{Fellow,~IEEE}
    \thanks{This work was partially funded by the Samsung Advanced Institute of Technology, the Swiss National Science Foundation HEAR-EAR, 200021\_172553 and the SNSF Sinergia WeCare CRSII5177255 grants. (Corresponding author: Chang Gao)}
    \thanks{Chang Gao, Xi Chen, Shih-Chii Liu, Tobi Delbruck are with the Institute of Neuroinformatics, University of Z\"urich and ETH Z\"urich, Z\"urich, Switzerland {\tt\small chang@ini.uzh.ch, xi@ini.uzh.ch, shih@ini.uzh.ch, tobi@ini.uzh.ch}}% <-this % stops a space
    \thanks{Antonio Rios-Navarro is with the Robotic and Technology of Computers Lab, Universidad de Sevilla, Seville, Spain {\tt\small arios@us.es}}% <-this % stops a space
    % \thanks{Manuscript received April 19, 2005; revised J 26, 2020.}
}
% \markboth{IEEE JOURNAL ON EMERGING AND SELECTED TOPICS IN CIRCUITS AND SYSTEMS,~Vol.~XX, No.~X, June~2020}%
% {Shell \MakeLowercase{\textit{et al.}}: Bare Demo of IEEEtran.cls for IEEE Journals}


% \IEEEspecialpapernotice{(Invited Paper)}

\IEEEaftertitletext{\vspace{-2\baselineskip}}
\maketitle

\begin{abstract}
Low-latency, low-power portable recurrent neural network (RNN) accelerators offer powerful inference capabilities for real-time applications such as IoT, robotics, and human-machine interaction. 
We propose a lightweight Gated Recurrent Unit (GRU)-based RNN accelerator called EdgeDRNN that is optimized for low-latency edge RNN inference with batch size of 1. 
EdgeDRNN adopts the spiking neural network inspired delta network algorithm to 
exploit temporal sparsity in RNNs. 
Weights are stored in inexpensive DRAM which enables EdgeDRNN to compute large multi-layer RNNs on the most inexpensive FPGA.
The sparse updates reduce DRAM weight memory access by a factor of up to 10x
and the delta can be varied dynamically to trade-off between latency and accuracy. 
EdgeDRNN updates a 5 million parameter 2-layer GRU-RNN in about 0.5\,ms. 
It achieves latency comparable with a 92\,W Nvidia 1080 GPU.
It outperforms NVIDIA Jetson Nano, 
Jetson TX2 and Intel Neural Compute Stick 2 in latency by 5X. 
For a batch size of 1, 
EdgeDRNN achieves a mean effective 
throughput of 20.2\,GOp/s and a wall plug power efficiency that is over 4X higher than the commercial edge AI platforms. 
\end{abstract}


% \fi
\IEEEpeerreviewmaketitle

%-----------------------------------------------------------------------------------------------------------
\section{Introduction}
\label{sec:introduction}
\thispagestyle{arXiv}
Deep neural networks (DNNs) have been widely applied to solve various practical problems with state-of-the-art performance. 
Recurrent neural networks (\textbf{RNN}s) which are a subset architecture of DNNs, are particularly useful in applications involving time series inputs, such as speech recognition~\cite{graves2012,Graves2013} and dynamical system control\cite{funahashi_approximation_1993,chow_recurrent_1998}. In contrast to Convolutional Neural Networks (\textbf{CNN}s) which use filter kernels, 
RNNs are fully-connected networks: They take a 1D vector as input and produce a vector of output. 
The feature vectors generated by CNNs can be fed into an RNN for further processing. 
In this way, RNNs can connect the high dimensional input features over time, 
which is useful for complex sequential classification or regression tasks.
Gated RNNs modify a ``vanilla'' RNN to add nonlinear operations to the units that allow them to memorize and gate their output.
Long Short-Term Memory units (\textbf{LSTM})~\cite{lstm_hoch97} 
and Gated-Recurrent Units (\textbf{GRU})~\cite{gru_og} are used to  
overcome the vanishing gradient problem frequently encountered during vanilla RNN 
training with backpropagation through time (\textbf{BPTT}), 
where the sequential operations of the RNN are unrolled to compute the weight updates based on output error.
By using BPTT with labeled training data, GRU and LSTM RNNs can be trained to high accuracy for tasks involving time series such as continuous speech recognition.
\begin{figure}[!t]
	\centering
    \includegraphics[width=.8\linewidth]{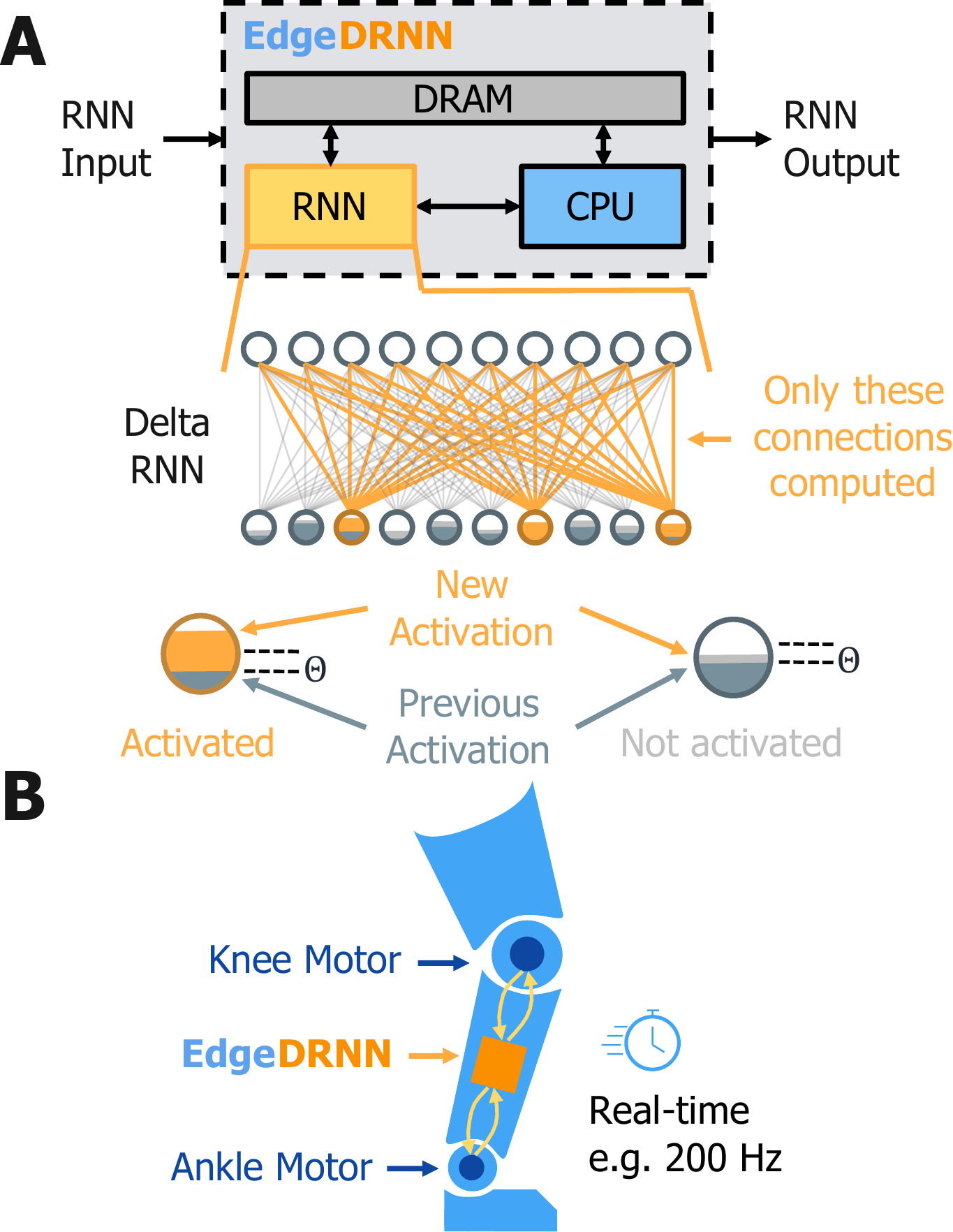}
    \caption{A: EdgeDRNN accelerator concept. B: Example target application from~\cite{ampro}.}
    \label{fig:edgedrnn-concept}
\end{figure}

Edge computing devices that embed some intelligence implemented through trained DNNs are gaining interest in recent years. 
An advantage of edge computing is that computations are done locally on end-user devices to reduce latency and protect privacy~\cite{chen2019}.
Most literature reports on the use of CNNs for edge devices. There is less reported on devices that use RNNs particularly on embedded low-latency, 
high energy-efficient platforms that use FPGAs. 
RNNs also have larger memory footprints and memory access of the fully-connected weight matrices dominates power consumption.
RNNs are usually computed on the cloud which introduces large and variable latency,
thereby making it hard to guarantee real-time performance for edge applications such as human-computer interaction devices and mobile,
robotic applications. 

Optimization methods have been applied to RNNs for embedded hardware implementations (e.g. weight pruning in ESE~\cite{han2017ese}, and structure pruning in BBS~\cite{bbs2019}). 
We also previously reported the DeltaRNN accelerator~\cite{deltarnn} that uses the delta network algorithm~\cite{neil2016delta}. 
Our first DeltaRNN implementation~\cite{deltarnn} stored the large weight matrices 
in FPGA block RAM and thus needed expensive FPGA boards with greater than 15\,W power 
consumption.
However, this work only focused on pushing the limit of high batch-1 throughput without considering the memory and power constraints of extreme edge devices.
Typical chips for edge applications, such as microcontrollers and small-footprint FPGA, only have a few hundreds of kilobytes (kB) of on-chip SRAM, but large RNNs usually have megabytes (MB) of parameters, making them difficult to be buffered on-chip even after compression. 
In this case, %despite introducing off-chip memory bottleneck, 
storing RNN parameters in off-chip memory such as flash or DDR memory is an inevitable choice for edge devices.
Therefore, reported hardware RNN implementations~\cite{han2017ese,bbs2019} 
cannot be easily scaled down for edge platforms.

This work is on EdgeDRNN, a hardware accelerator for RNN computation on the edge~\cite{gaoAICAS2020}. 
Because of our interest in real-time edge applications, 
our focus is on supporting low-latency batch-1 inference of large RNNs for real-time performance on low-cost and low power but very constrained edge platforms.
We show that our implementation can run large-scale multi-layer RNNs using a small number of processing elements with the large weight matrices stored on off-chip DRAM. 
Besides accelerating RNN inference, it leaves most cycles of the CPU in the system-on-chip (\textbf{SoC}) available for other tasks, such as feature extraction and I/O management. 
EdgeDRNN can be implemented on a small-footprint FPGA with 19X less logic cells and 15X less on-chip memory compared to the one used in DeltaRNN~\cite{deltarnn}.
Thus, EdgeDRNN is suitable for embedded system applications such as robotics (Fig.~\ref{fig:edgedrnn-concept}B).

Moreover, in previous work using the current delta network algorithm, a global threshold is applied on both the inputs and hidden unit activations of every layer of the network in sparsifying the activation vector. In this work, we looked at how different threshold values on the inputs and activations of hidden units affect the trade-off between the accuracy of the network on a regression task and the sparsity levels of the change in the activation vectors. A higher sparsity level implies reduced weight memory access and reduced computes.

This paper makes the following contributions:
\begin{enumerate}
    \item We describe a flexible, low-cost, high throughput edge FPGA RNN accelerator that uses a spiking neural network-inspired Delta RNN principle to provide state-of-art latency and power efficiency for a wide range of gated RNN network sizes with batch size of 1.
    \item We report the first study of a delta network that uses different delta thresholds for the input and activations of the hidden units. On our tested regression task, this modification increases temporal sparsity of hidden delta states by 16\% compared to using a global threshold.
    \item We compare the usability and throughput performance of two different EdgeDRNN implementation on the SoC FPGA: Bare-metal and embedded Linux.
    The latter enables faster development and we report the correct FPGA memory bus port configuration that minimizes the performance loss due to CPU contention for the memory controller.
    \item We report benchmark latency and throughput numbers of RNN inference on state-of-the-art commercial chips for edge applications. To our best knowledge, these numbers have never been reported before. 
    
        % \item To enable application and improvement, we will open the design of EdgeDRNN and PyTorch code to train delta networks at \url{https://sites.google.com/view/EdgeDRNN} under a license that allows free non-commercial use.
    % \item This paper describes the first RNN accelerator for edge applications. 
    % \item EdgeDRNN is the first RNN accelerator that exploits temporal sparsity in conjunction with large weight matrices stored in DRAM. The accelerator exploits temporal sparsity using the delta network (\textbf{DeltaGRU})~\cite{neil2016delta} algorithm. It achieves sub-millisecond inference of big multi-layer RNNs comparable with a desktop-level GPU, but with 38 times less power.
\end{enumerate}

The rest of this paper is organized as follows.
Section~\ref{sec:background} describes the background of gated recurrent unit (GRU) based RNN and the algorithm of delta network based GRU-RNN, which is called \textsl{DeltaGRU}. 
Section~\ref{sec:EdgeDRNN} describes the architectural design of the accelerator and its implementation on MiniZed.
Section~\ref{sec:results} discusses experimental results including the experiments using different delta thresholds for the network. 
Section~\ref{sec:compare} compares the proposed accelerator with prior work and commercial edge devices. 
Section~\ref{sec:conclude} concludes the paper. 

%---------------------------------------------------------------------------------------------
\section{Background}
\label{sec:background}
%---------------------------------------------------------------------------------------------
%---------------------------------------------------------------------------------------------
\subsection{ DNN hardware optimization methods:}
Various methods have been proposed to reduce the dominant RNN matrix-vector (\textbf{MxV}) operations.
Static approximation methods (i.e. constructed during training) include \textit{quantization}, \textit{arithmetic}, and \textit{weight pruning}.

\textbf{Quantization:} Quantizing floating-point weights or activations to fixed-point numbers with shorter bit width reduces memory footprint of networks and make it possible to use fixed-point MAC units instead of expensive floating-point MAC units~\cite{Chang2017, Lee2016, Guan2017FCCM, han2017ese}. 
Chip area can be further reduced by replacing conventional fixed-point multipliers by look-up table based~\cite{Shin2017} or multiplexer~\cite{yangbnn2018} based multipliers on low bit precision networks with 2-4 bit weights. 
By including quantization during training (e.g. by using an approach like dual-copy rounding~\cite{stromatias2015robustness}) it is possible to reduce weight
precision to 8 bits without accuracy loss.

\textbf{Weight pruning:} Pruning removes unimportant neuron connections that results in sparse weight matrices~\cite{Han2015}. 
Sparse matrices can be encoded into a sparse matrix format such as the Compressed Sparse Column (\textbf{CSC}) and Compressed Sparse Row (\textbf{CSR}). 
With an accelerator that can decode the sparse matrix format on-chip, the sparse matrix-vector (\textbf{SpMV}) multiplication can be accelerated by executing multiply-and-accumulate (\textbf{MAC}) operations only on nonzero weights. This approach was adopted by the Efficient Speech Recognition Engine (\textbf{ESE})~\cite{han2017ese}. 
Because unstructured pruning results in computation that is hard to balance across processing elements, 
structured pruning methods have also been proposed to improve load balancing during the SpMV computation~\cite{bbs2019,Kadetotad2020}.
This approach was used by the LSTM FPGA accelerator using Bank Balanced Sparsity (\textbf{BBS})~\cite{bbs2019}. It is also used by the custom digital IC of~\cite{Kadetotad2020} where it is called Hierarchical Coarse Grain Sparsity (HCGS).
Structured pruning is a popular approach for improving RNN hardware performance; both BBS and HCGS use it to 
increase effective MAC efficiency, but large increases in efficiency result in significantly worse inference accuracy~\cite{Kadetotad2020}. For example, a 16X compression increases the error rate by a factor of about 1.2X. 
It allows static (compile-time) optimization, but training is fairly complicated since exploration of the additional structure hyperparameter values is needed %need to be carefully explored 
to find optimum values that are matched to the particular hardware.

\textbf{Arithmetic:} 
Bit-serial NN accelerators such as~\cite{Judd2016, Bilaniuk2019} utilize a flexible bit-serial MAC to support various precision of network parameters and are smaller in area compared to conventional fixed-point MAC units. 
However, since a bit-serial MAC units requires more cycles to finish a multiplication between high bit precision operands, more bit-serial MAC units are required to achieve higher throughput than using conventional MAC units and larger adder trees are need for accumulating partial sums. Thus, the average speedup using this method is only around 2X but it comes with extra overhead area.
The C-LSTM accelerator used Toeplitz-like weight matrices in the form of blocked circulant matrices to reduce RNN memory requirements~\cite{Wang2017} since multiple rows in each circulant matrix block can be generated from a single vector. 
The method also enables the use of Fast Fourier Transform (\textbf{FFT}) to reduce the MxV cost from $\mathcal{O}(n^2)$ to $\mathcal{O}(n\log(n))$~\cite{Wang2018}.
However, forcing weight matrices to be blocked circulant is coarse-grained and leads to higher accuracy degradation compared to weight pruning~\cite{han2017ese, bbs2019}. Moreover, the method leads to hardware overhead of computing the FFTs of activations and weights.

\textbf{Temporal sparsity:} 
The delta network algorithm~\cite{neil2016delta} capitalizes on the temporal sparsity of activation state vectors in a network.
Setting a finite threshold that is greater than zero has the effect of zeroing-out below-threshold elements of the activation vector, which results in sparse delta vectors.
Since zero activations have no downstream influence, these MACs can be skipped. 
It means that entire columns of the weight matrix can be skipped. 
Thus delta networks marry the temporal sparsity of spiking networks with
the synchronous update and analog state transmission of conventional deep networks. Combining
these principles provides the 
benefits of sparse computing and efficient communication of precise analog 
information with reduced and predictable memory access of inexpensive DRAM which is crucial for storing the weights.
%
%It might seem like skipping neuron updates inevitably reduces inference accuracy. But 

A set of studies~\cite{neil2016delta,gaoAICAS2020, ampro} 
showed in a variety of networks that by applying the delta principle during training, the accuracy loss is minimal even with a 5-10X improvement of RNN throughput and latency.
For example, \cite{neil2016delta} used a 4-layer 320 units per layer GRU RNN for continuous speech recognition on the Wall Street Journal dataset. The Word Error Rate (\textbf{WER}) increased by only a factor of 1.08X but with a reduced memory access of 6.2X.

\subsection{Gated Recurrent Unit}

The update equations for a GRU layer of $H$ neurons and $I$-dimensional input, which are as follows:
\begin{equation}
\begin{aligned}
\mathbf{r}_{t} &=\sigma\left(\mathbf{W}_{xr}\mathbf{x}_{t}+\mathbf{W}_{hr}\mathbf{h}_{t-1}+\mathbf{b}_{r}\right)\\
\mathbf{u}_{t} &=\sigma\left(\mathbf{W}_{xu}\mathbf{x}_{t}+\mathbf{W}_{hu}\mathbf{h}_{t-1}+\mathbf{b}_{u}\right)\\
\mathbf{c}_{t} &=\textrm{tanh}\left(\mathbf{W}_{xc}\mathbf{x}_{t}+\mathbf{r}_{t}\odot\left(\mathbf{W}_{hc}\mathbf{h}_{t-1}\right)+\mathbf{b}_{c}\right)\\
\mathbf{h}_{t} &=\left(1-\mathbf{u}_{t}\right)\odot \mathbf{c}_{t}+\mathbf{u}_{t}\odot \mathbf{h}_{t-1}
\label{eq:1}
\end{aligned}
\end{equation}

\noindent where $\mathbf{r}, \mathbf{u}, \mathbf{c}\in\mathcal{R}^{H}$ are the \textit{reset gate}, 
the \textit{update gate} and the \textit{cell state} respectively. $\mathbf{W}_{x}\in\mathcal{R}^{H\times I}$, $\mathbf{W}_{h}\in\mathcal{R}^{H\times H}$
are weight matrices and $\mathbf{b}\in\mathcal{R}^{H}$ are bias vectors. 
The $\sigma$ variable denotes the logistic sigmoid function. 
Each GRU input is the vector $\mathbf{x}_{t}$ and its output is the vector $\mathbf{h}_t$.

Fig.~\ref{fig:gru}A illustrates the update of the normal GRU reset gate as a flow diagram.

\subsection{DeltaGRU}
\begin{figure*}[t]
    \begin{minipage}{1.0\textwidth}
      \centering
      \includegraphics[width=0.98\textwidth]{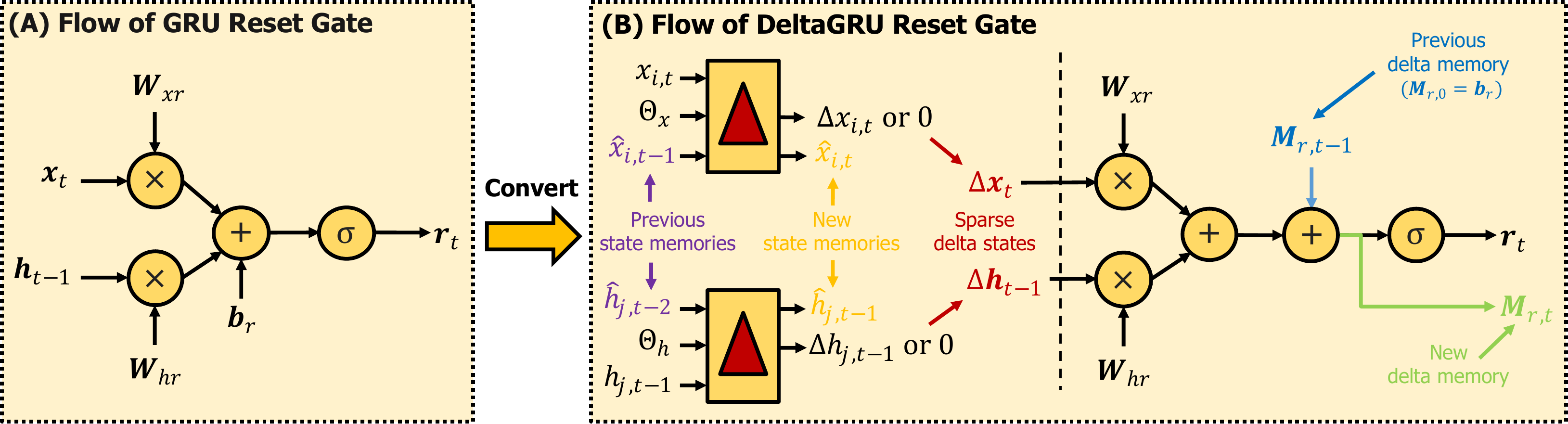}
    \end{minipage}%
	\centering
    \caption{Computation flow of the reset gate in gated recurrent units (GRUs). A: Normal GRU reset gate update. B: A DeltaGRU reset gate update. See Eqs.\ref{eq:1}-\ref{eq:3}.}
    \label{fig:gru}
\end{figure*}

The delta network method is applied to the GRU-RNN architecture; we call this \textit{DeltaGRU}. 
Assume an input vector sequence $\mathbf{X}=\{\mathbf{x}_{t}\mid t=1,2,...,T\}$ with sequence length of $T$, we first declare the following variables: 
%To transform a normal GRU-RNN to a DeltaGRU, we first define the following terms:
\begin{equation}
\begin{aligned}
\hat{x}_{i, t} &= \left\{\begin{matrix}
    x_{i, t} & ,| x_{i, t} - \hat{x}_{i, t-1} | \geq \Theta_{x} \\
    \hat{x}_{i, t-1} & , | x_{i, t} - \hat{x}_{i, t-1} | < \Theta_{x}
	\end{matrix}\right. \\
\hat{h}_{j, t} &= \left\{\begin{matrix}
    h_{j, t} & ,| h_{j, t} - \hat{h}_{j, t-1} | \geq \Theta_{h} \\
    \hat{h}_{j, t-1} & , | h_{j, t} - \hat{h}_{j, t-1} | < \Theta_{h} 
	\end{matrix}\right. \\
\Delta x_{i, t} &= \left\{\begin{matrix}
	x_{i,t} - \hat{x}_{i,t-1}&, | x_{i,t} - \hat{x}_{i, t-1} | \geq \Theta_{x}\\ 
	0 & ,| x_{i,t} - \hat{x}_{i, t-1} | < \Theta_{x}
    \end{matrix}\right. \\
\Delta h_{j, t} &= \left\{\begin{matrix}
	h_{j,t} - \hat{h}_{j,t-1}&, | h_{j,t} - \hat{h}_{j,t-1} | \geq \Theta_{h}\\ 
	0 & ,| h_{j,t} - \hat{h}_{j,t-1} | < \Theta_{h}
    \end{matrix}\right.
\end{aligned}
\label{eq:2}
\end{equation}
where $\hat{x}_{i, t}$ is the $i$-th element of input state memory vectors $\hat{\mathbf{x}}_{t}$ in timestep $t$. 
$\hat{h}_{j, t}$ is the $j$-th element of hidden state memory vectors $\hat{\mathbf{h}}_{t}$ in timestep $t$. 
$\Delta x_{i, t}$ is the $i$-th element of delta input state vectors $\Delta \mathbf{x}_{t}$. 
$\Delta h_{j, t}$ is the $j$-th element of delta hidden state vectors $\Delta \mathbf{h}_{t}$.
$\Theta_{x}$ and $\Theta_{h}$ are respectively the delta thresholds of inputs and hidden state for each layer. In the initial timestep ($t=1$), $\hat{x}_{i, 0}$, $h_{i, 0}$, $\hat{h}_{i, -1}$ are all initialized to zeros.

The update equations for the DeltaGRU are now:
\begin{equation}
\begin{aligned}
	\mathbf{M}_{r,t} &= \mathbf{W}_{xr}\Delta \mathbf{x}_{t} + \mathbf{W}_{hr}\Delta \mathbf{h}_{t-1} + \mathbf{M}_{r, t-1} \\
	\mathbf{M}_{u,t} &= \mathbf{W}_{xu}\Delta \mathbf{x}_{t} + \mathbf{W}_{hu}\Delta \mathbf{h}_{t-1} + \mathbf{M}_{u, t-1} \\
	\mathbf{M}_{xc, t} &= \mathbf{W}_{xc}\Delta \mathbf{x}_{t}+ \mathbf{M}_{xc, t-1} \\
	\mathbf{M}_{hc, t} &= \mathbf{W}_{hc}\Delta \mathbf{h}_{t-1} + \mathbf{M}_{hc, t-1} \\
	\mathbf{r}_{t} &= \sigma(\mathbf{M}_{r,t}) \\
	\mathbf{u}_{t} &= \sigma(\mathbf{M}_{u,t}) \\
	\mathbf{c}_{t} &= \tanh(\mathbf{M}_{xc, t} + \mathbf{r}_{t}\odot \mathbf{M}_{hc, t}) \\
	\mathbf{h}_{t} &= (1-\mathbf{u}_{t})\odot \mathbf{c}_{t} + \mathbf{u}_{t} \odot \mathbf{h}_{t-1}
\end{aligned}
\label{eq:3}
\end{equation}
where $\mathbf{M}_{r,t=0} = \mathbf{b}_{r}$, $\mathbf{M}_{u, t=0} = \mathbf{b}_{u}$, $\mathbf{M}_{xc, t=0} = \mathbf{b}_{c}$, $\mathbf{M}_{hc, t=0} = 0$ are delta memory vectors and $\mathrm{M}\in\mathcal{R}^{H}$. 
Variables $\sigma$ and $\odot$ indicate the sigmoid function and element-wise multiplication of vectors respectively.

Fig.~\ref{fig:gru}B illustrates these operations for the DeltaGRU reset gate. 
The input vector $\mathbf{x}_{t}$ and the hidden state vector $\mathbf{h}_{t-1}$ are respectively replaced by the delta input state vector $\Delta \mathbf{x}_{t}$ and the delta hidden state vector $\Delta \mathbf{h}_{t-1}$. 
Values of the previous state memory vectors $\hat{\mathbf{x}}_{t-1}$, $\hat{\mathbf{h}}_{t-2}$ are updated using Eq.~\ref{eq:2} to generate new state memory vectors $\hat{\mathbf{x}}_{t}$, $\hat{\mathbf{h}}_{t-1}$. 
The previous delta memory vector $\mathbf{M}_{r,t-1}$ holds the previous step's partial sum-product and the resulting new delta memory vector $\mathbf{M}_{r,t}$ is stored. Otherwise the operations are the same as for the original GRU reset gate, 
as shown in Fig.~\ref{fig:gru}A. The other gates have similar flow diagrams. The state and delta memories are 1D vectors and can be easily fit into on-chip SRAM buffers.

%-----------------------------------------------------------------------------------------
\subsection{Temporal Sparsity}
The temporal sparsity $\mathbf{\Gamma}$ of a DeltaGRU network of $L$ layers with an input sequence length of $T$ is defined as the fraction of zeros in the $\Delta_{x}$ and $\Delta_{h}$ vectors,
signified by $\Gamma_{\Delta x}$ and $\Gamma_{\Delta h}$ respectively. 
The effective temporal sparsity $\Gamma_{\rm Eff}$ is the weighted average of $\Gamma_{\Delta x}$ and $\Gamma_{\Delta h}$ according to the number of network parameters they correspond to. 
The definition of temporal sparsity is given by Eq.~\ref{eq:4}:
\begin{equation}
\begin{aligned}
	\Gamma_{\Delta x} &= \frac{1}{L\cdot T \cdot I}\sum_{t=1}^{T}n^{1}_{x, t} + \frac{1}{(L-1)\cdot T \cdot H}\sum_{l=2}^{L}\sum_{t=1}^{T}n^{l}_{x, t}  \\
	\Gamma_{\Delta h} &= \frac{1}{L\cdot T \cdot H}\sum_{l=1}^{L}\sum_{t=1}^{T}n^{l}_{h, t} \\
	\Gamma_{\rm Eff} &= \frac{\left(3HI+3H^2(L-1)\right)\cdot\Gamma_{\Delta x}+3H^2L\cdot\Gamma_{\Delta h}}{3HI+3H^2(L-1)+3H^2L} \\
	&= \frac{\left(I+H(L-1)\right)\cdot\Gamma_{\Delta x}+HL\cdot\Gamma_{\Delta h}}{I+H(L-1)+HL}
\end{aligned}
\label{eq:4}
\end{equation}
where $n^{l}_{x, t}$ and $n^{l}_{h, t}$ are the number of zero elements respectively in the delta vectors $\Delta x$ and $\Delta h$ in layer $l$ at timestep $t$. 
Because operations on biases are negligible, they are ignored in Eq.~\ref{eq:4}.

By skipping zero elements in delta vectors, 
whole columns of matrix-vector MAC operations 
can be skipped. If the delta network is properly trained (by including the delta operation), \cite{neil2016delta,ampro,GaoISCAS2019} showed that the number of operations can be reduced by 5X to 100X with negligible loss of accuracy, depending on the temporal evolution of the states of the input and hidden units.

%-----------------------------------------------------------------------------------------
\subsection{Datasets}
Two datasets are used in this paper: the \textsl{TIDIGITS}~\cite{leonard1993tidigits} dataset for the classification task and the 
\textsl{SensorsGas} dataset for the regression task.
The \textsl{TIDIGITS} speech dataset has more than 25k digit sequences spoken by over 300 men, women, and children. 
The entire training and test sets are used in our experiments.
The \textsl{SensorsGas} dataset consists of recordings of metal-oxide sensors in response to various concentrations of carbon monoxide gas over 14 days~\cite{burgues2018estimation, burgues2018multivariate}. 
This dataset was used in~\cite{sensorsgas} to evaluate the network performance of a gated RNN in predicting the concentration of carbon monoxide. 
The dataset used here comes from the 70/30 Split variant, that is, 
70\% of the sequences are randomly selected to form the training set, 
while the remaining sequences form the test set.

%*********************************************************************************************
\section{EdgeDRNN Accelerator}
\label{sec:EdgeDRNN}
%*********************************************************************************************
%---------------------------------------------------------------------------------------------
\subsection{Overview}
%---------------------------------------------------------------------------------------------
\begin{figure}[!t]
	\centering
    \includegraphics[width=0.98\linewidth]{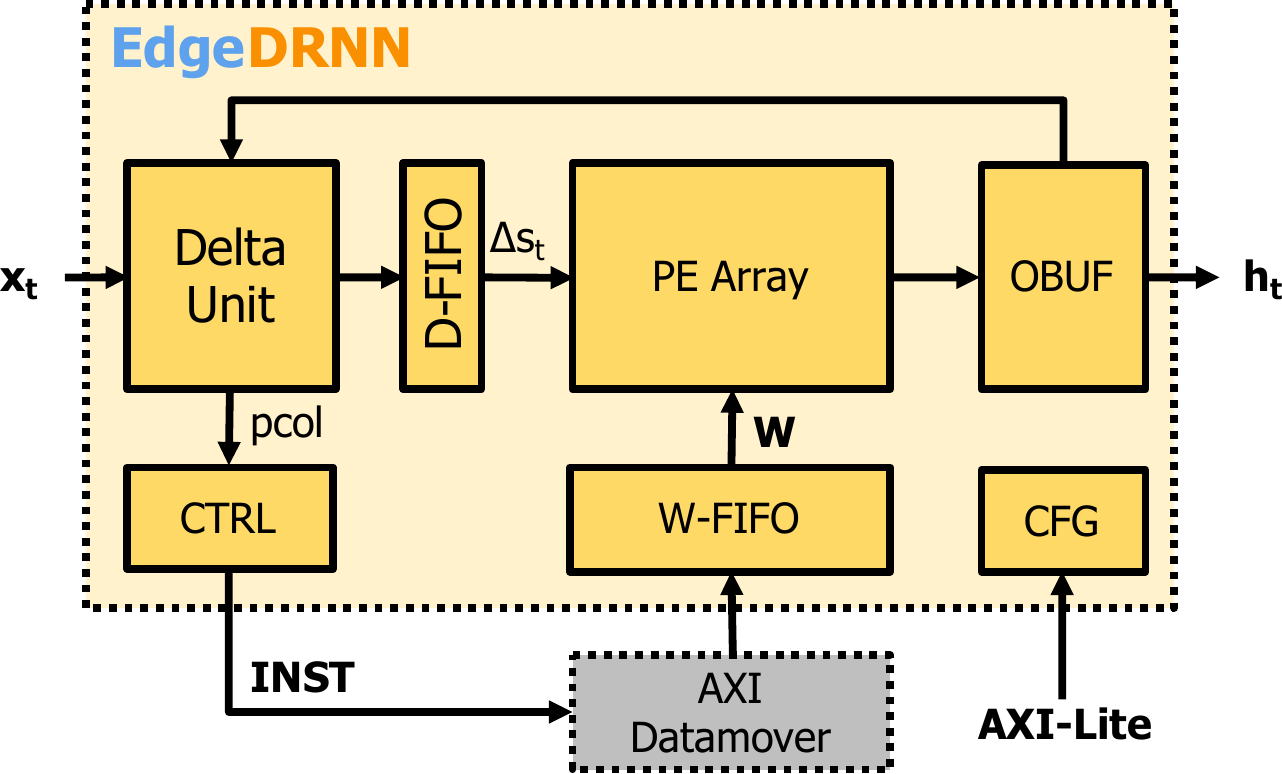}
    \caption{EdgeDRNN accelerator architecture}
    \label{fig:EdgeDRNN}
\end{figure}

Due to limited weight reuse, it is difficult to compute RNNs efficiently for real-time applications that usually work best with a batch size of 1.
% \footnote{We can run EdgeDRNN in batch mode by instantiating multiple EdgeDRNN blocks on the PL. 
% Batch processing can be useful in some real-time applications, e.g. when using a multi-shooting prediction approach.}
Therefore, a big challenge of RNN inference on the edge is the scarce off-chip memory bandwidth available on portable platforms, 
and the limited amount of on-chip block RAM on small FPGAs. 
EdgeDRNN uses cheap off-chip DRAM for weight storage and reduces memory bandwidth by exploiting temporal sparsity in RNN updates. 

Fig.~\ref{fig:EdgeDRNN} shows the architecture of the EdgeDRNN accelerator.
The main modules consist of the Delta Unit for encoding delta vectors and generating weight column pointers (\textbf{pcol}); 
the Processing Element (\textbf{PE}) Array for matrix-sparse vector multiplications; 
the  (\textbf{CTRL}) control module which contains finite state machines (\textbf{FSMs}) and encodes instructions to control the AXI Datamover. 
Other modules include the configuration module (\textbf{CFG}) composed of configuration registers; 
the output buffer (\textbf{OBUF}) for buffering and redirecting outputs back to the Delta Unit and the \textbf{W-FIFO} for buffering weights.

%---------------------------------------------------------------------------------------------
\subsection{Delta Unit \& CTRL}
%---------------------------------------------------------------------------------------------
\begin{figure}[!t]
	\centering
    \includegraphics[width=0.98\linewidth]{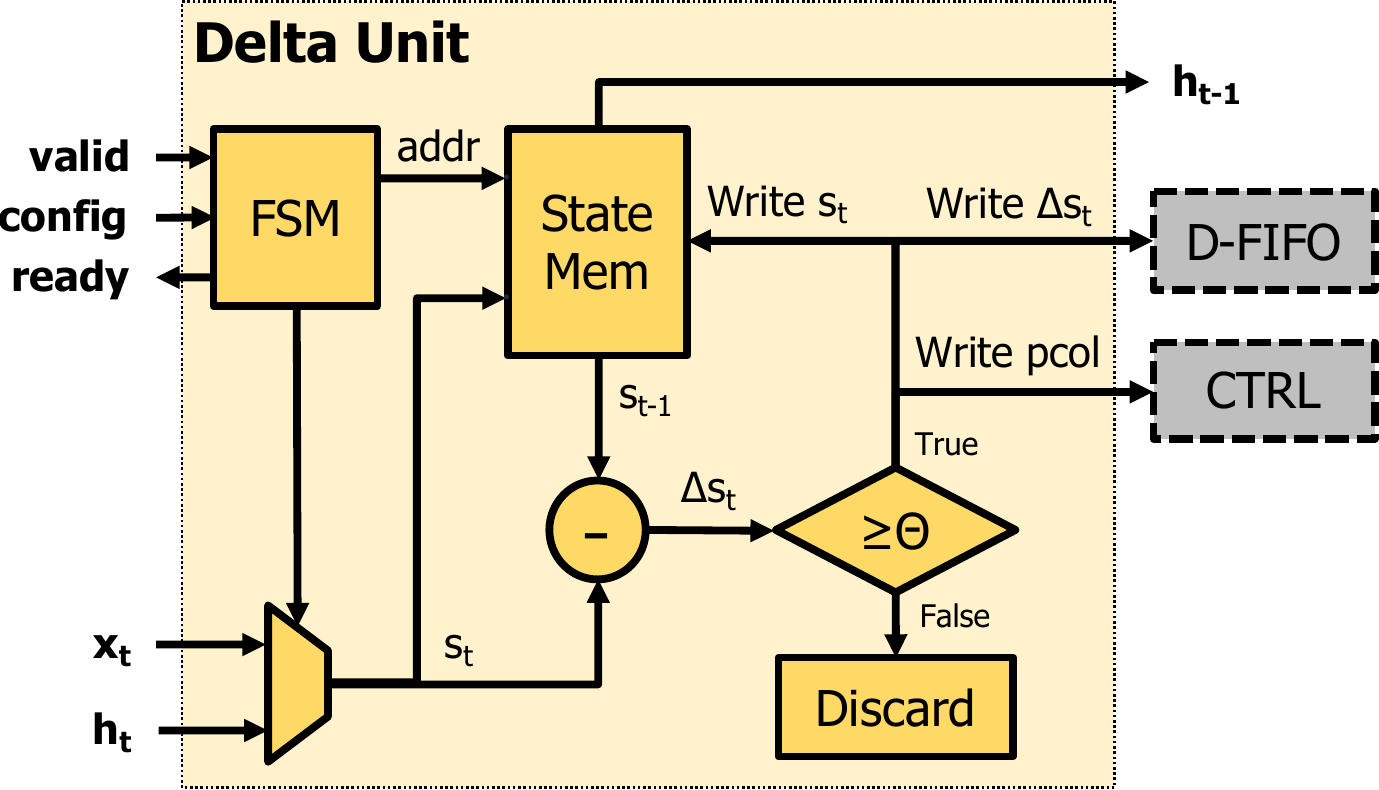}
     \label{fig:delta_unit}
\caption{Flow chart of the Delta Unit.}
\end{figure}

The Delta Unit stores state memory for delta state vector encoding in a block random access memory
(\textbf{BRAM}\footnote{BRAM is the standard SRAM memory block on FPGAs; on Xilinx Zynq FPGAs a single BRAM block has 18-bit words and a capacity of 18kb}). 
The FSM addresses the BRAM according to the \texttt{valid} signal of input state vectors $x_{t}$ and $h_{t}$,
one of which, is selected to be processed as $s_{t}$ at a time depending on the FSM state. 
The Delta Unit encodes one element of a delta state vector $\Delta s_{t}$ in each clock cycle after the \texttt{valid} signal asserted until the whole vector is processed. 

The vector sizes are provided by the \texttt{config} signal from the CFG module. 
Delta state vector elements that are greater than or equal to threshold $\Theta_{x}$ or $\Theta_{h}$; and their corresponding physical weight column address pointer (\texttt{pcol}) are respectively dispatched to the D-FIFO and CTRL. 
The corresponding state element $s_{t}$ is written into the BRAM to update the state memory. Otherwise, elements are discarded without being written into the D-FIFO. 
By using only $1$ Delta unit, the latency in clock cycles for the Delta Unit to process a vector is exactly the length of that vector. 
It is possible to reduce the latency by searching for nonzero elements in $N$ subsections of a vector simultaneously. 
It can be realized by using $N$ Delta Unit blocks in parallel to fill at most one nonzero value into the D-FIFO on every clock cycle. 
Assuming that nonzero elements are uniformly distributed in a delta state vector and using $N$ Delta Unit blocks running in parallel, the latency $\tau_{DU}$ in clock cycles to process a whole vector is
\begin{equation}
	\tau_{DU} \approx \max\left(\left\lceil \frac{D}{N\cdot d}\right\rceil, \lceil D \cdot\left(1-\Gamma\right)\rceil\right)
\end{equation}
where $D$ is the length of the vector, $d$ is the length of the subsection of the vector or the look-ahead window side of the Delta Unit; 
%$N$ is the number of Delta Units computing in parallel, 
and $\Gamma$ is the temporal sparsity defined in Eq.~\ref{eq:4}.

Although $\tau_{DU}$ can be hidden under $\tau_{m}$, 
	the latency of computing MxV, $\tau_{DU}$ becomes a bottleneck of total latency when $\tau_{DU} > \tau_{m}$, which could happen when an accelerator uses a large number of MAC units to compute small networks.
However, in this work, we aim to run large network inference with a small number of MAC units for edge applications, making $\tau_{DU}\ll \tau_{m}$; 
thus, $\eta=1$ is used in EdgeDRNN. 
The MAC utilization results shown in Section IV.D prove that this choice did not lead to latency bottleneck.

The CTRL module contains FSMs that control the PE array. This module generates 80-bit instructions for controlling the Xilinx AXI Datamover IP~\cite{axi_datamover} to fetch RNN parameters. 
The instruction contains \texttt{pcol} and the burst length calculated from the dimensions of the network stored in configuration registers. 

%---------------------------------------------------------------------------------------------
\subsection{Processing Element Array}
\begin{figure}[!t]
	\centering
	\includegraphics[width=0.98\linewidth]{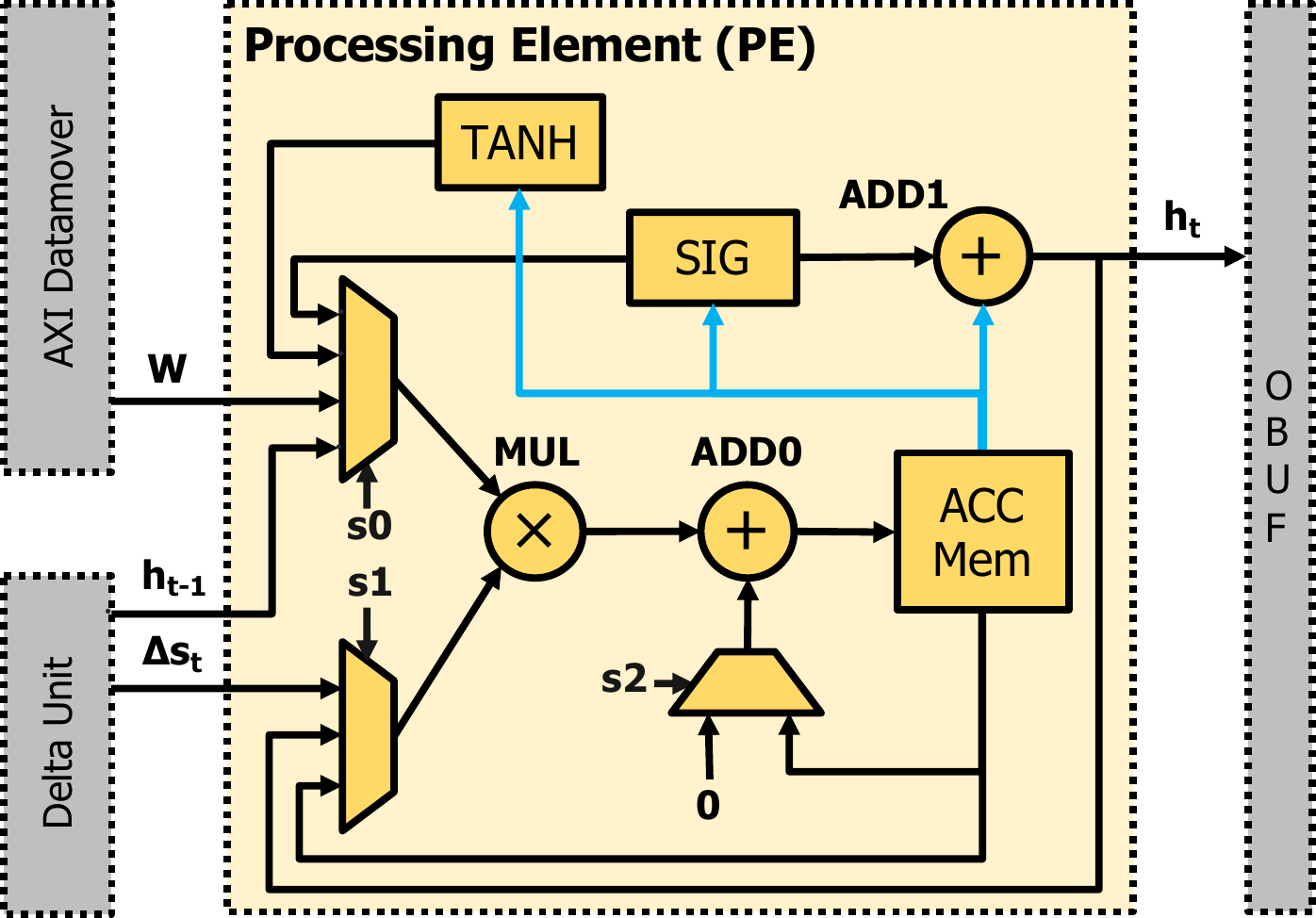}
	\caption{Architecture of the EdgeDRNN processing element (PE).}
	\label{fig:pe}
\end{figure}

%---------------------------------------------------------------------------------------------
Two-dimensional arithmetic unit arrays such as systolic arrays are difficult to be fully utilized in portable edge devices due to 
scarce on-chip memory resources, the low external memory bandwidth of the system and the limited weight reuse nature of RNNs.
In order to fully utilize every PE, a vector PE array is used in EdgeDRNN. Fig.~\ref{fig:pe} show the internal structure of a PE.

The PE has a 16-bit multiplier \textbf{MUL} and two adders, 
32-bit \textbf{ADD0} and 16-bit \textbf{ADD1}.
Multiplexers are placed before operands of MUL so that the PE can be reused for
both MxV and vector dot products. 
The multiplexer below ADD0 selects between '0' and the BRAM data. '0' is chosen when an initialization of BRAM is needed as shown in Fig.~\ref{fig:pe}. 
ADD1 is responsible for element-wise vector additions.
All units are parameterized in the System Verilog RTL and configurable at compile-time to support any fixed-point precision within their designed bit width.
The PE supports \texttt{tanh} and \texttt{sigmoid} functions by using look-up tables (\textbf{LUTs}). The input bit width of LUTs is fixed to 16 bits while the output bit width can be set anywhere between 5 (Q1.4) to 9 (Q1.8) bits.
\begin{figure}[!t]
	\centering
	\includegraphics[width=0.85\linewidth]{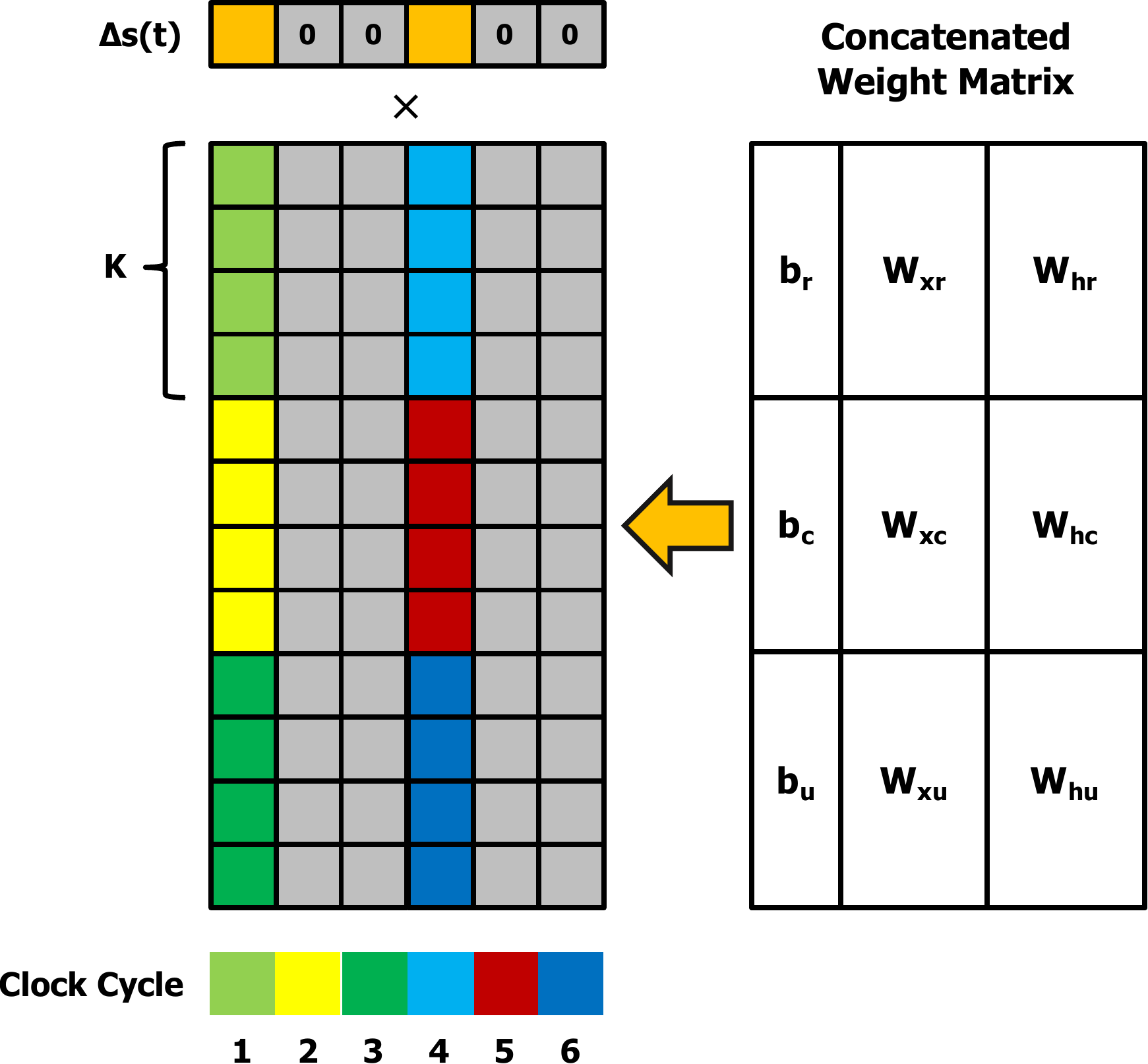}
	\caption{Flow of sparse matrix-vector multiplication in EdgeDRNN and the arrangement of GRU weights in the concatenated weight matrix.}
     \label{fig:mxv}
\end{figure}

Fig.~\ref{fig:mxv} shows the sparse MxV flow. %process of MxV. 
The weight matrices of the GRU-RNN are concatenated following the arrangement shown on the right half of the figure. 
Biases are appended to the concatenated weight matrix as the first column and an element $1$ is appended to each input state vector $\mathbf{x}_t$ as the first element. 
The PE array multiplies only nonzero delta state elements with corresponding valid columns. 
Products are accumulated in the Accumulation Memory (\textbf{ACC Mem}) to compute delta memory vectors $\mathbf{M}_{r}$, $\mathbf{M}_{u}$, $\mathbf{M}_{ic}$, $\mathbf{M}_{hc}$. 
Products involving $\mathbf{b}_{r}$, $\mathbf{W}_{ir}$, $\mathbf{W}_{hr}$ are accumulated to $\mathbf{M}_{r}$; 
$\mathbf{b}_{u}$, $\mathbf{W}_{iu}$, $\mathbf{W}_{hu}$ to $\mathbf{M}_{u}$; 
$\mathbf{b}_{c}$, $\mathbf{W}_{ic}$ to $\mathbf{M}_{ic}$; 
$\mathbf{W}_{hc}$ to $\mathbf{M}_{hc}$.
According to the delta update scheme defined by Eq.~\ref{eq:2}, 
the appended $1$ in the delta state vector $\Delta \mathbf{x}_{t}$ becomes $0$ after the initial timestep, 
which means that biases $\mathbf{b}_{r}$, $\mathbf{b}_{u}$, $\mathbf{b}_{c}$ are only accumulated to the ACC Mem by once and will be skipped by the Delta Unit after the initial timestep.

\begin{figure*}[!t]
	\centering
	\includegraphics[width=0.8\linewidth]{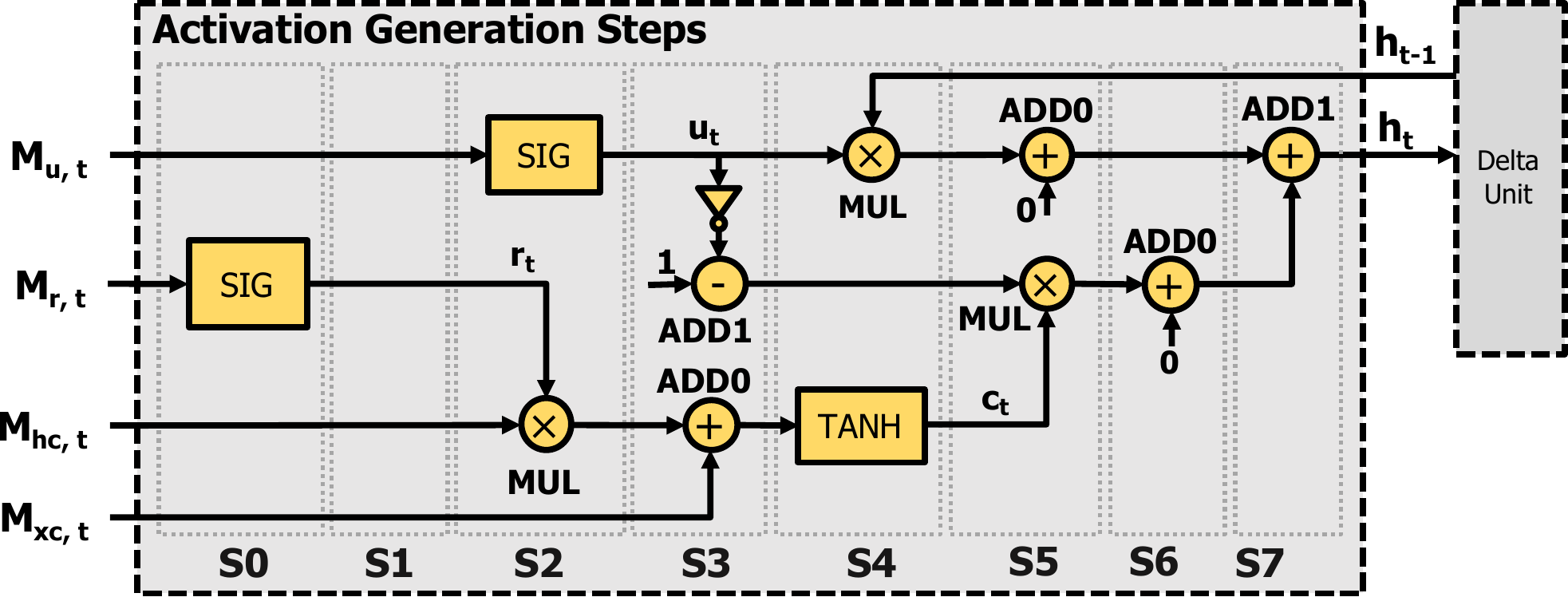}
	\caption{Stages of GRU activation pipeline in the PE Array.}
	\label{fig:act}
\end{figure*}
The calculation of activation $h_{t}$ after the MxV is also done by the PE array and stages of this process are shown in Fig.~\ref{fig:act}. 
The PE array fetches the delta memory vectors from the ACC Mem to calculate $h_{t}$ in 8 pipeline stages. 
Paths without any operator in any stage are buffered for 1 clock cycle using flip-flops. 
During execution of the activation $h_t$ generation, 
stages \textbf{S0$\sim$S2} are executed simultaneously with \textbf{S5$\sim$S7} to reuse the arithmetic units using time-division multiplexing.

Finally, assuming that the DRAM interface can deliver $W_{\rm DRAM}$ bits per RNN clock cycle for weight fetch, the optimum number $K$ of PEs in the array is determined by the weight precision bit width $W_{\rm Weight}$. The definition of $K$ and corresponding theoretical peak throughput, $\nu_{Peak}$, is defined below:
\begin{equation}
\begin{aligned}
    K          &= W_{\rm DRAM}/W_{\rm Weight}, \\
    \nu_{Peak} &= 2\cdot f_{pl}\cdot K
\end{aligned}
\label{eq:6}
\end{equation}
where $f_{pl}$ is the clock frequency of the programmable logic. For example, the FPGA used in this paper has a 64-bit DRAM interface, so, with 16-bit weights, $K=8$ is optimal.

%-----------------------------------------------------------------------
\subsection{Implementation on MiniZed}
\begin{figure}[!t]
	\centering
	\includegraphics[width=0.98\linewidth]{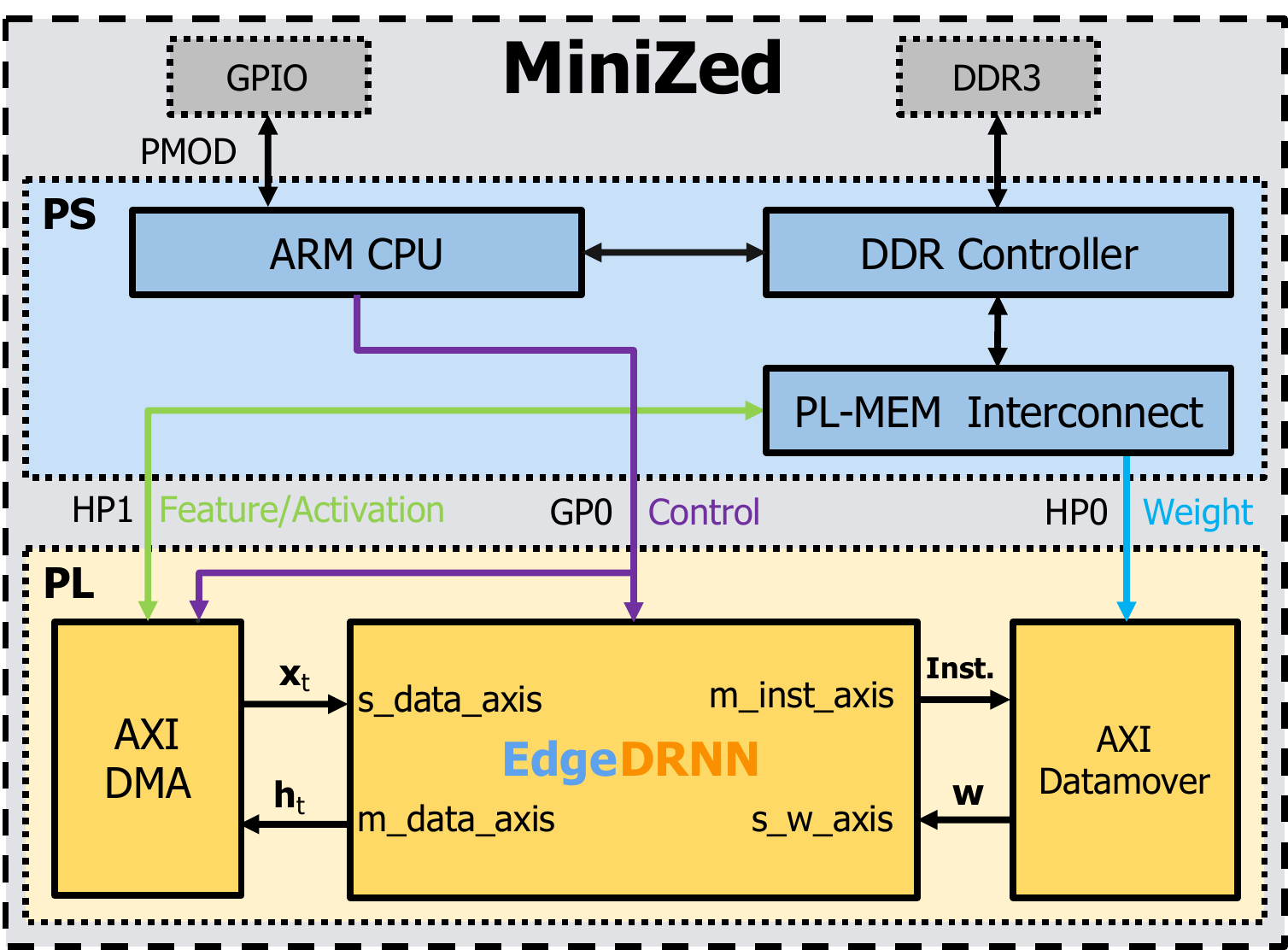}
	\caption{Top-level diagram of the EdgeDRNN implementation on the MiniZed development board.}
	\label{fig:top}
\end{figure}
Fig.~\ref{fig:top} shows the latest implementation of EdgeDRNN on the \$89  
MiniZed development board~\cite{minized} which has a Zynq-7007S SoC.
EdgeDRNN is implemented in the programmable logic (\textbf{PL}). 
The SoC also has a programmable CPU which is in a module called the Processing System (\textbf{PS}). 
Data is transferred between PS and PL through high performance (\textbf{HP}) slave ports while control signals generated by the PS is transferred through general purpose (\textbf{GP}) master ports.
The hard IP block, AXI Datamover, is controlled by the PS to fetch weights to the PL from DDR3L memory. Another hard IP block, AXI DMA is controlled by the PS to transfer inputs and outputs of the accelerator.
Compared to our previous work~\cite{gaoAICAS2020}, we reduced resource utilization by replacing the AXI SmartConnect IP with the AXI Interconnect IP while preserving the same throughput and latency. 
To further reduce on-chip power, we used the power optimization strategy during implementation in Xilinx Vivado and lower the ARM CPU clock frequency from 667 MHz to 400 MHz for the bare-metal version.

The peak DRAM read memory bandwidth is 1 GB/s at the 125 MHz clock frequency (64-bits$\times$125~MHz/8-bits/byte). 
EdgeDRNN can be configured to support 1, 2, 4, 8, 16-bit fixed-point weights and 16-bit fixed-point activations. In this paper, EdgeDRNN is configured to support 16-bit activation and 8-bit weights.
To fully exploit this HP port bandwidth, we implement $K=8$ PEs following Eq.~\ref{eq:6}.
Adding more PEs would only leave them idle since weight fetches are limited by the DRAM bandwidth.

The AXI-Lite General Purpose (\textbf{GP}) master port is used for the single-core 
ARM Cortex-A9 CPU to control the AXI-DMA and to write the configuration to the accelerator.
Configurations include physical start address of the concatenated weights, delta thresholds, and network dimensions.

\begin{table}[!t]
\caption{Resource utilization of MiniZed using 5-bit (Q1.4) LUT}
\label{tab:resource}
\centering
\resizebox{0.5\textwidth}{!}{ % RESIZEBOX
\begin{tabular}{|l|c|c|c|c|c|}
\hline
\textbf{}           & \textbf{LUT} & \textbf{LUTRAM} & \textbf{FF} & \textbf{BRAM} & \textbf{DSP} \\ \hline
\textbf{Available}  & 14400        & 6000            & 28800       & 50            & 66           \\ \hline
\textbf{EdgeDRNN}   & 30.8\%       & 0.4\%           & 9.3\%       & 32\%          & 13.6\%      \\ \hline
\textbf{Total}      & 65.2\%       & 4.4\%           & 34.1\%      & 66\%          & 13.6\%      \\ \hline
\end{tabular}
}
\end{table}
The PL is driven by single clock domain of 125\,MHz generated by the PS. 
Table~\ref{tab:resource} shows the resource utilization solely for EdgeDRNN (with 5-bit (Q1.4) LUTs) and for the whole PL after synthesis and implementation. 
BRAMs are used to store previous state memory in the Delta Units and the
accumulation memory in PEs and FIFOs. 
8 DSPs are used for the MAC units 
in the 8 PEs while the remaining DSP in CTRL produces weight column addresses.
The most consumed resources are LUTs (72\%). 
This entry-level XC7Z007S FPGA has only 14.4k LUTs. 
By comparison, the top level XC7Z100 has 19X more LUTs and 11X more BRAM.

\subsection{Petalinux OS Integration}
Xilinx's Zynq chips are hosted on heterogeneous embedded platforms with a variety of peripherals and communication interfaces. To work with this type of system there are two workflows, bare-metal and embedded OS. 

The bare-metal workflow is similar to the workflow of conventional microcontrollers. Bare-metal has a set of libraries that establish a very thin software layer over all the hardware resources available in the system and that helps a little during the elaboration of the software that will be deployed in the system; however, detailed knowledge of the hardware is still necessary to ensure correct functionality. The resulting software runs on the PS processor making use of all its computing power since it is the only software running on the core. Bare-metal allows a more dedicated use of the system resources to achieve high performance execution but it offers little flexibility and versatility.

The second option is to use an embedded Linux OS provided by Xilinx called PetaLinux. This OS establishes several software layers over the system hardware that simplifies its use and the development of applications that make use of the system's peripherals like USB, Bluetooth, and Wi-Fi. The Linux system is a preemptive multitasking operating system that can make application development much faster. Since running Linux slightly slows down inference (Sec.~\ref{sec:results}), users can decide to pay the throughput price of using Linux for faster development time and easier maintenance. For EdgeDRNN, we implemented both systems to meet our various application requirements.

%*********************************************************************************************
\section{Experimental Results}
\label{sec:results}
We previously developed two EdgeDRNN system-level demonstrations: continuous spoken digit recognition~\cite{GaoISCAS2019} and real-time control of a powered leg prosthetic~\cite{ampro}. Here we report the results of new experiments to measure accuracy, throughput, and power efficiency on the spoken digit task and a new regression task on gas concentration estimation. We also report measurements of embedded Linux implementation of EdgeDRNN.
%---------------------------------------------------------------------------------------------
\subsection{Experimental Setup: Training}
We evaluate the accuracy of DeltaGRU and the hardware performance of EdgeDRNN using this DeltaGRU network on both a classification task using the \textsl{TIDIGITS}~\cite{leonard1993tidigits} dataset and on a regression task using the \textsl{SensorsGas}~\cite{sensorsgas} dataset.
\subsubsection{Classification}
For the classification task, we trained 6 different DeltaGRU network sizes and compared their WER on the \textsl{TIDIGITS} audio digit dataset, 
evaluated using the greedy decoder. 
Inputs to the networks consist of 40-dimensional log filter bank features extracted from audio sampled at 20\,kHz using a frame size of 25\,ms and frame stride of 10\,ms. 
We use the Connectionist Temporal Classification (\textbf{CTC}) loss~\cite{Graves2006} to handle variable input sequence lengths. The DeltaGRU networks were trained for 50 epochs using a learning rate of 3e-4 and batch size of 32. Following a similar procedure in~\cite{GaoISCAS2019}, a continuous spoken digit recognition demo is built using EdgeDRNN to prove the system functionality\footnote{\url{https://www.youtube.com/watch?v=XyN-jh5yiMI}}.

\subsubsection{Regression}
For the \textsl{SensorsGas} regression task, the input dimension of the network is 14 corresponding to data from the 14 sensors. 
We adopt a 2-step pretrain and retrain scheme we developed for~\cite{ampro}: 
1) We pretrain a \texttt{cuDNN} GRU model on the training set for 100 epochs. 
The learning rate is 5e-4 and the batch size of 64. 
2) We load these parameters into a DeltaGRU network with same size as the \texttt{cuDNN} GRU and retrain for another 10 epochs with learning rate of 3e-3 and batch size of 256. 
In this step we optimize the deltas for the visible and hidden units. 
Because the cuDNN GRU model is highly optimized for NVIDIA GPUs, 
the pretrain step helps to train the network to achieve high accuracy with 5X less time. 

All networks are trained using the \texttt{Adam} optimizer and quantization-aware training using quantization scheme similar to~\cite{stromatias2015robustness}. 
To improve accuracy,  we use nonlinear functions with the same input and output bit precision as the LUT in the forward phase of the training. In the backward phase, the gradient of the nonlinear function is calculated using the original nonlinear functions in FP32 precision. 
Training was done with PyTorch 1.2.0 on NVIDIA GPUs running CUDA\,10 and cuDNN\,7.6.
\begin{figure}[!t]
	\centering
	\includegraphics[width=0.48\textwidth]{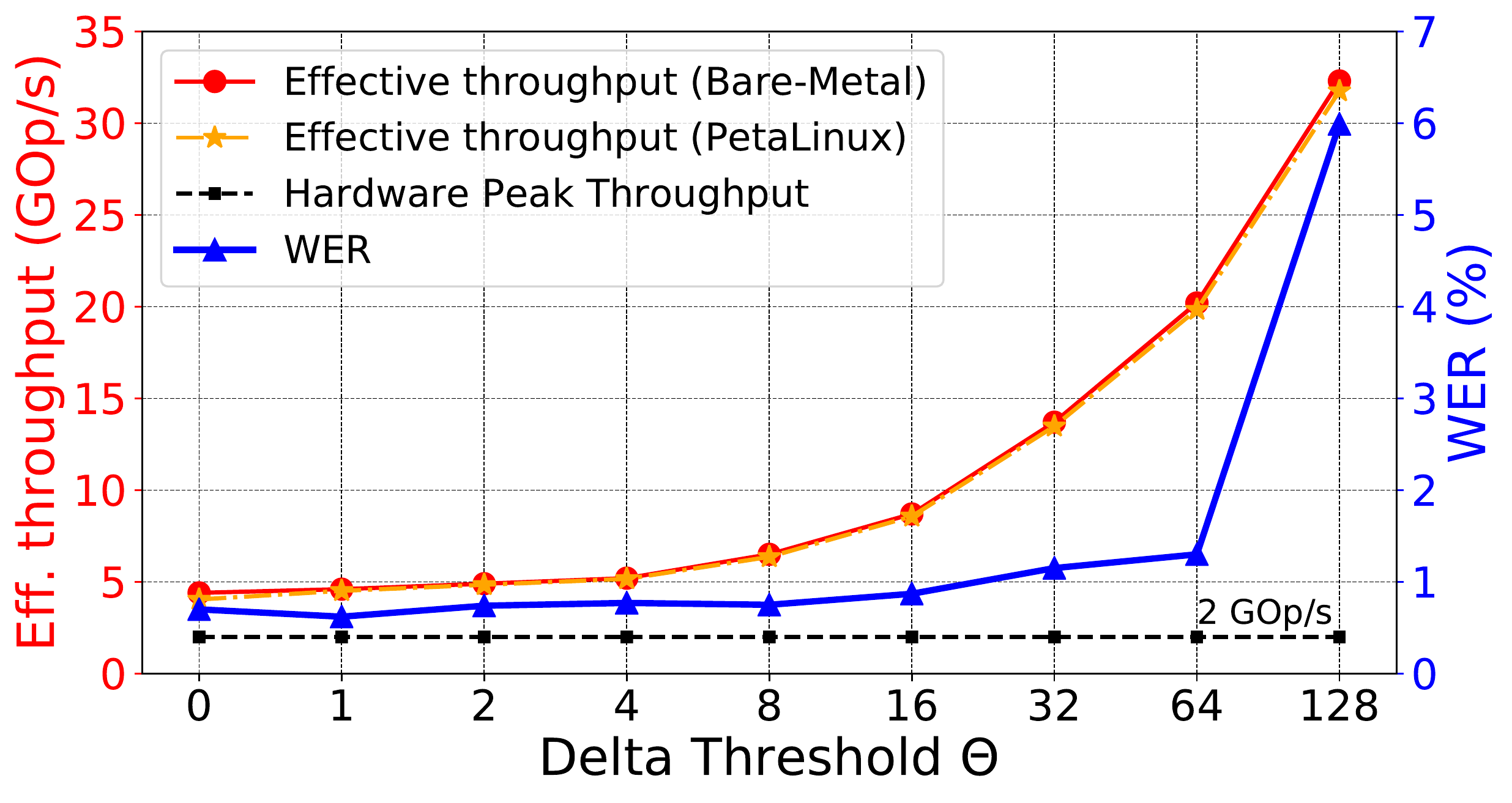}
	\caption{Mean effective throughput and word error rate evaluated on the \textsl{TIDIGITS} test set as a function of the delta threshold used in both training and inference of a 2L-768H-DeltaGRU network.  $\Theta_{x} = \Theta_{h}$ ($\Theta$ is shown as Q8.8 integer values corresponding to 0$\sim$0.5 floating point threshold).}
	\label{fig:delta}
\end{figure}
%----------------------------------------------------------------------
\subsection{Experimental Setup: Network Implementation}

After the quantized DeltaGRU is trained for a particular task, a Python script converts the PyTorch network modules into C/C++ hardware network header files. These files contain the network parameters and configuration register values for EdgeDRNN. 
By including the header files, bare-metal or PetaLinux applications are compiled using the standard cross compiler. 
The resulting system image is transferred to the QSPI flash (bare-metal) or eMMC storage (PetaLinux) on the MiniZed. 
In each timestep of the RNN, a feature vector is transferred from the PS to the accelerator using the AXI DMA. For measuring the performance of the accelerator, features are calculated offline on a computer and stored in a header file. For using the accelerator in real-world applications, features such as log filter bank and spike count features for audio, are calculated online by the ARM core in the PS. A flag connected to a PS hardware register is raised at the end of each timestep.

%---------------------------------------------------------------------------------------------
\subsection{Accuracy \& Throughput}
%--------------------------------------------------------------------------------------------
\subsubsection{Classification}

Fig.~\ref{fig:delta} shows the EdgeDRNN throughput and WER on the \textsl{TIDIGITS} test set versus the $\Theta$ 
used in training and testing of a 2L-768H-DeltaGRU network. $\Theta$ is the same for both $\Delta x$ and $\Delta h$.
With $K=8$\,PEs and PL frequency $f_{pl}=125$\,MHz, 
EdgeDRNN has a theoretical peak throughput of $2Kf_{pl}=2$\,GOp/s.
At $\Theta=0$, there is still a speedup of about 2X from natural sparsity of the delta vectors.
Higher $\Theta$ leads to better effective throughput, but with gradually increasing WER.
The optimal point is at $\Theta=64$ (0.25), just before a dramatic increase of WER,
where EdgeDRNN achieves an effective throughput around 20.2\,GOp/s with 1.3\% WER. 
WER and throughput of smaller tested networks are shown in Table~\ref{tab:benchmark_baremetal}.
The 5-bit (Q1.4) LUT was used for this task and did not lead to accuracy loss compared to the network running on CPU with FP32 nonlinear functions.
%--------------------------------------------------------------------------------------------
\begin{table*}[]
\caption{Word error rate, latency and throughput of EdgeDRNN running with Bare-Metal on DeltaGRU networks trained with $\Theta=64$, $\beta=$1e-5.}
\label{tab:benchmark_baremetal}
\centering
\resizebox{1.00\textwidth}{!}
{
\begin{tabular}{|c|c|c|c|c|c|c|c|c|c|c|c|c|}
\hline
\multirow{2}{*}{\textbf{Network Sizes}} & \multirow{2}{*}{\textbf{\begin{tabular}[c]{@{}c@{}}Op\\ (Timestep)\end{tabular}}} & \multirow{2}{*}{\textbf{WER}} & \multirow{2}{*}{\textbf{\begin{tabular}[c]{@{}c@{}}WER Degradation\\vs. GRU\end{tabular}}} & \multicolumn{4}{c|}{\textbf{Latency ($\mu$s)}}              & \multicolumn{2}{c|}{\textbf{Eff. Throughput (GOp/s)}} & \multirow{2}{*}{\textbf{\begin{tabular}[c]{@{}c@{}}MAC \\ Utilization\end{tabular}}} & \multirow{2}{*}{\textbf{$\Gamma_{\Delta x}$}} & \multirow{2}{*}{\textbf{$\Gamma_{\Delta h}$}} \\ \cline{5-10}
                                        &                                                                                   &                               &                                                                                             & \textbf{Min} & \textbf{Max} & \textbf{Mean} & \textbf{Est.} & \textbf{Mean}           & \textbf{Est.}          &                                                                                      &                                               &                                               \\ \hline
1L-256H                                 & 0.5 M                                                                             & 1.83\%                        & +1.36\%                                                                                     & 16.7         & 142.2        & 46.2          & 43.3          & 9.9                     & 10.5                   & 495\%                                                                                & 25.6\%                                        & 90.0\%                                        \\ \hline
2L-256H                                 & 1.2 M                                                                             & 1.13\%                        & +0.69\%                                                                                     & 29.1         & 258.9        & 90.7          & 91.6          & 13.7                    & 13.6                   & 685\%                                                                                & 78.9\%                                        & 89.1\%                                        \\ \hline
1L-512H                                 & 1.7 M                                                                             & 1.04\%                        & +0.44\%                                                                                     & 40.6         & 331.2        & 130.6         & 129.8         & 13.0                    & 13.1                   & 650\%                                                                                & 25.6\%                                        & 89.5\%                                        \\ \hline
2L-512H                                 & 4.9 M                                                                             & 0.89\%                        & +0.75\%                                                                                     & 57.3         & 656.8        & 252.6         & 262.9         & 19.2                    & 18.4                   & 960\%                                                                                & 85.5\%                                        & 91.2\%                                        \\ \hline
1L-768H                                 & 3.7 M                                                                             & 1.27\%                        & +0.11\%                                                                                     & 64.1         & 616.7        & 224.3         & 224.8         & 16.6                    & 16.6                   & 830\%                                                                                & 25.6\%                                        & 91.3\%                                        \\ \hline
2L-768H                                 & 10.8 M                                                                            & 0.77\%                        & +0.53\%                                                                                     & 96.5         & 1344.5       & 535.6         & 541.6         & 20.2                    & 19.9                   & 1010\%                                                                               & 87.0\%                                        & 91.6\%                                        \\ \hline
\end{tabular}
}
\end{table*}
\begin{table*}[!t]
\caption{Latency and throughput of EdgeDRNN running with PetaLinux on DeltaGRU networks trained with $\Theta=64$, $\beta=$1e-5.}
\label{tab:benchmark_petalinux}
\centering
\begin{tabular}{|c|c|c|c|c|c|c|}
\hline
\multirow{2}{*}{\textbf{Network Sizes}} & \multirow{2}{*}{\textbf{\begin{tabular}[c]{@{}c@{}}Op\\ (Timestep)\end{tabular}}} & \multicolumn{3}{c|}{\textbf{Latency ($\mu$s)}}                   & \textbf{Eff. Throughput (GOp/s)} & \multirow{2}{*}{\textbf{\begin{tabular}[c]{@{}c@{}}MAC \\ Utilization\end{tabular}}} \\ \cline{3-6}
                                        &                                                                                   & \multicolumn{1}{l|}{\textbf{Min}} & \textbf{Max} & \textbf{Mean} & \textbf{Mean}                    &                                                                                      \\ \hline
1L-256H                                 & 0.5 M                                                                             & 17.0                              & 311.0        & 48.2          & 9.5                              & 475\%                                                                                \\ \hline
2L-256H                                 & 1.2 M                                                                             & 30.0                              & 461.0        & 93.1          & 13.4                             & 670\%                                                                                \\ \hline
1L-512H                                 & 1.7 M                                                                             & 42.0                              & 603.0        & 133.6         & 12.7                             & 635\%                                                                                \\ \hline
2L-512H                                 & 3.7 M                                                                             & 59.0                              & 923.0        & 257.5         & 18.8                             & 940\%                                                                                \\ \hline
1L-768H                                 & 4.8 M                                                                             & 66.0                              & 627.0        & 228.5         & 16.3                             & 815\%                                                                                \\ \hline
2L-768H                                 & 10.8 M                                                                            & 99.0                              & 1366.0       & 544.9         & 19.8                             & 990\%                                                                                \\ \hline
\end{tabular}
\end{table*}
%--------------------------------------------------------------------------------------------
\subsubsection{Regression}
\begin{figure}[!t]
	\centering
	\includegraphics[width=0.48\textwidth]{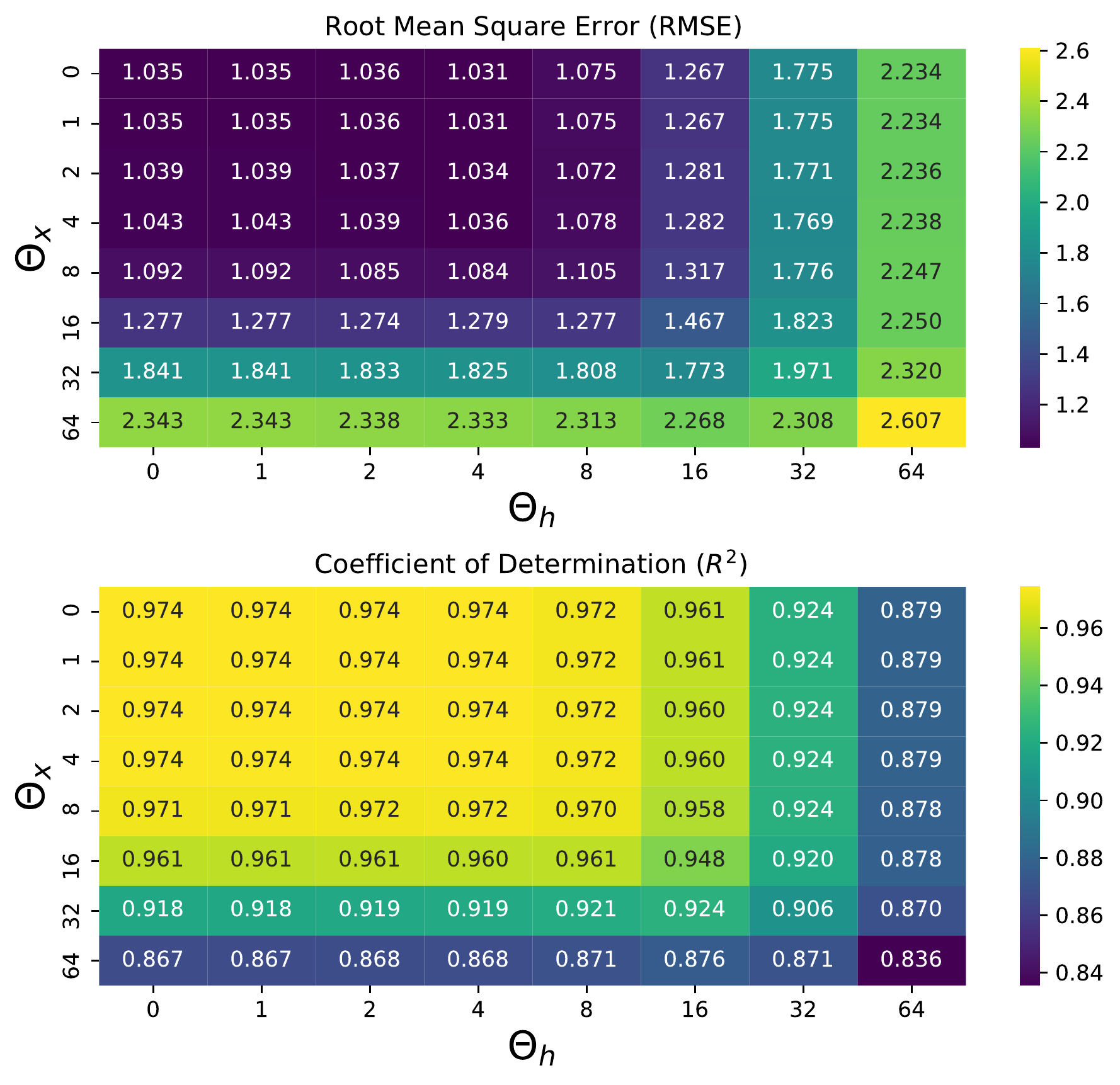}
	\caption{RMSE (smaller is better) and $R^2$ (larger is better) versus  $\Theta_x$ and $\Theta_h$ of the 2L-256H-DeltaGRU model evaluated on the \textsl{SensorsGas} test set.}
	\label{fig:error}
\end{figure}

\begin{figure}[!t]
	\centering
	\includegraphics[width=0.48\textwidth]{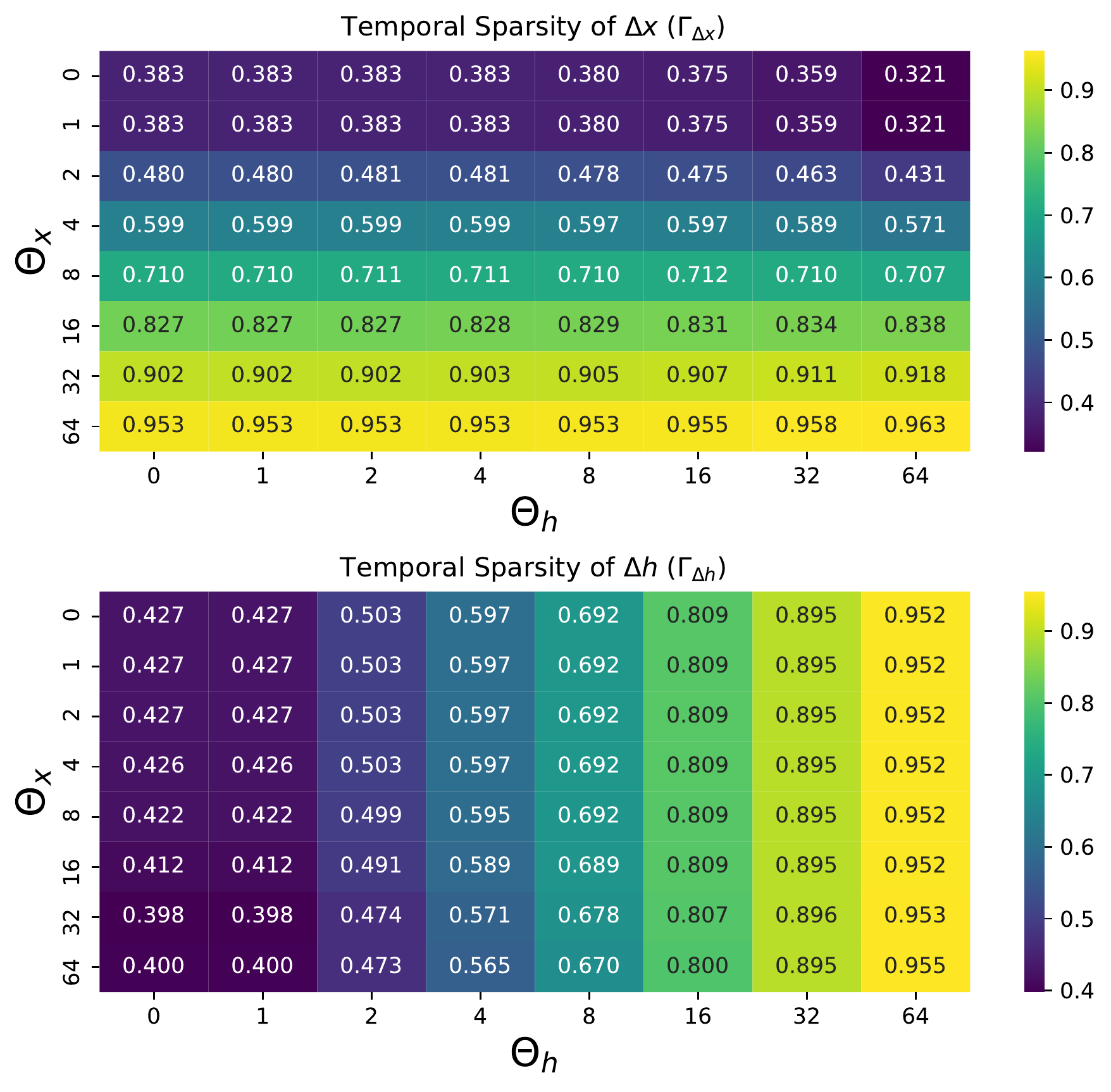}
	\caption{Temporal sparsity $\Gamma_{\Delta x}$ and $\Gamma_{\Delta h}$ versus $\Theta_x$ and $\Theta_h$ of the 2L-256H-DeltaGRU model evaluated on the \textsl{SensorsGas} test set.}
	\label{fig:sp}
\end{figure}

In this regression task, we evaluate the impact of using different delta thresholds for $\Delta x$ and $\Delta h$ on the accuracy results of a 2L-256H-DeltaGRU model evaluated on the \textsl{SensorsGas} testset.
Fig.~\ref{fig:error} and Fig.~\ref{fig:sp} show respectively the regression accuracy and temporal sparsity versus $\Theta_x$ and $\Theta_h$.
%of a 2L-256H-DeltaGRU model evaluated on the \textsl{SensorsGas} test set. 
The pretrained 2L-256H-GRU network without using a delta threshold, achieves a root-mean-square error (\textbf{RMSE}) of 0.995 and coefficient of determination ($R^2$) of 0.976. 
This accuracy was achieved using 5-bit (Q1.4) LUTs, which gave the lowest RMSE out of all other LUT bit precision values.
% Thus, the gas concentration is very highly predictable from the input sensors.

Similar to the results for the classification task, Fig.~\ref{fig:error} shows that 
the accuracy degrades when larger delta thresholds are used. 
Fig.~\ref{fig:sp} shows that the sparsity levels of $\Gamma_{\Delta x}$ and $\Gamma_{\Delta h}$ are heavily influenced by their corresponding delta thresholds. 
The accuracy degrades faster with increasing $\Theta_x$ for a fixed $\Theta_h$ than with increasing $\Theta_h$ for a fixed $\Theta_x$.
$\Theta_x$ has a minor impact on $\Gamma_{\Delta_h}$ and vice versa. 
The results from this regression task indicate that propagating changes more often in input states is more important than propagating changes in hidden states.
By exploiting this phenomenon, we get the optimal point $(\Theta_x, \Theta_y)=(4, 8)$, where the RMSE and $R^2$ are 1.078 and 0.972 respectively. 
With $\Gamma_{\Delta x}=59.7\%$ and $\Gamma_{\Delta h}=69.2\%$, 
the latency of the optimal model is 206\,$\mathbf{\mu}$s. 
In comparison, Jetson TX2 runs a 4.8X smaller 1L-200H-GRU network in 271\,$\mu$s~\cite{sensorsgas}.

%--------------------------------------------------------------------------------------------
\subsection{Theoretical \& Measured Performance}
%--------------------------------------------------------------------------------------------
Eq.~\ref{eq:8} gives the estimated mean effective throughput $\nu_{Eff}$ of EdgeDRNN running a DeltaGRU layer:
\begin{equation}
\begin{aligned}
\nu_{Eff}&=\frac{\mathrm{Op}}{\tau_{m}+\tau_{a}}
\\&\approx\frac{2\left(3HI+3H^2(L-1)+3H^2L\right)}{\frac{\left(3HI+3H^2(L-1)\right)(1-\Gamma_{\Delta x})+3H^2L(1-\Gamma_{\Delta h})}{Kf_{pl}}+\frac{3H}{Kf_{pl}}}
\end{aligned}
\label{eq:8}
\end{equation}
where $\mathrm{Op}$ is the number of operations 
in a DeltaGRU layer per timestep, $\tau_{m}$ the 
latency of MxV, $\tau_{a}$ the latency of remaining operations to produce the activation, and the other variables are defined as in Eqs.~\ref{eq:4} and~\ref{eq:6}. 

Table~\ref{tab:benchmark_baremetal} compares the Eq.~\ref{eq:8} predictions with benchmark results of different DeltaGRU network sizes running on EdgeDRNN. 
Estimated results calculated from Eq.~\ref{eq:8} are close to measured results and the maximum relative error between them is smaller than 7.1\%.  
Thus Eq.~\ref{eq:8} can be used to estimate EdgeDRNN performance for a particular RNN network size. 
On average, EdgeDRNN can run all tested networks
under 0.54\,ms latency corresponding to 20.2 GOp/s effective throughput for the 2L-768H-DeltaGRU.
%--------------------------------------------------------------------------------------------
\subsection{Performance in PetaLinux}
%--------------------------------------------------------------------------------------------
For performance measurements on the PetaLinux-based system, we implemented an application that performs the same operations as the software implemented for bare-metal but use the AXI DMA driver included in the OS.

Table \ref{tab:benchmark_petalinux} shows the latency and performance results for the 6 networks used in this work. 
The minimum and mean latency numbers in the PetaLinux version are respectively up to 3.4\% and 11.3\% higher than the numbers obtained for the bare-metal version. 
Because the minimum PetaLinux latency is nearly the same as the bare-metal latency, the big difference in maximum latency numbers between the PetaLinux and the bare-metal version is due to CPU contention for other tasks running on the PS that lock the single PS DDR controller. 
EdgeDRNN fetches weights from HP ports (Fig.~\ref{fig:top}) that are routed through the PS DDR controller. (The FPGA's ACP interface should not be used to access DRAM memory under PetaLinux because it is connected directly to the L2 cache on the ARM core where the OS runs. This configuration creates conflicts and the performance of the system is seriously compromised.) 
Under PetaLinux, the HP interface should be used to connect any module placed on the PL that requires direct access to the DRAM memory. 

To understand the impact of CPU load and CPU DRAM access on the RNN inference time, we wrote a small program that loops over a memory array and is designed to trigger L2 cache misses.
We used two different memory array sizes to study the effect of cache misses since the large memory array causes more L2 cache misses. 
Table~\ref{tab:cpudram} shows that the impact on RNN latency is minor: 
a small network takes about 50\% longer to run with either memory array size, 
and a large RNN is only slowed down by 10\%.

\begin{table}[!h]
\caption{EdgeDRNN RNN PetaLinux latency with CPU DRAM memory access.}
\centering
\begin{tabular}{l|c}
 &  Mean Latency (us)\\ \hline
\multicolumn{2}{c}{1L-256H-TH64 (little network with few parameters)}\\ \hline
Only EdgeDRNN & 48\\
EdgeDRNN + Small memory workload & 72\\
EdgeDRNN + Large memory workload & 82 \\ \hline
\multicolumn{2}{c}{2L-768H-TH64 (big network with many parameters)}\\ \hline
Only EdgeDRNN & 545\\
EdgeDRNN + Small memory workload & 613\\
EdgeDRNN + Large memory workload &  595 \\  \hline
\end{tabular}
\label{tab:cpudram}
\end{table}
The RNN inference time varies between 50\,us to 0.5\,ms across the different network sizes. 
During this inference time, the PS is free for other tasks (e.g. computing features) and only needs to check if the RNN update is finished when these tasks are completed.

%---------------------------------------------------------------------------------------------
\subsection{Power Measurement}
%---------------------------------------------------------------------------------------------
\begin{table}[!t]
\caption{Wall power breakdown of the MiniZed EdgeDRNN system during RNN inference (Bare-metal)}
\label{tab:power}
\centering
\begin{tabular}{|l|l|c|c|}
\hline
\multicolumn{2}{|c|}{\textbf{Part}}    & \textbf{Wall Power (mW)} & \textbf{Percentage} \\ \hline
\multirow{4}{*}{PS}   & DDR3L          & 534                      & 23.3\%              \\ \cline{2-4} 
                      & PLLs           & 388                      & 16.9\%              \\ \cline{2-4} 
                      & ARM Cortex-A9  & 166                      & 7.2\%              \\ \cline{2-4} 
                      & Peripherals    & 21                       & 0.9\%               \\ \hline
\multicolumn{2}{|l|}{Regulator/etc.}   & 942                      & 41.1\%              \\ \hline
\multicolumn{2}{|l|}{Static}           & 119                      & 5.2\%               \\ \hline
\multicolumn{2}{|l|}{\textbf{EdgeDRNN}}         & \textbf{66}                       & \textbf{2.9}\%               \\ \hline
\multicolumn{2}{|l|}{DMA/Interconnect} & 54                       & 2.4\%               \\ \hline
\multicolumn{2}{|l|}{Total}            & 2290                     &                     \\ \hline
\end{tabular}
\end{table}
Table~\ref{tab:power} shows the power breakdown of the MiniZed system. The total power is measured by a USB power meter; 
the PS, PL and static power is estimated by the Xilinx Power Analyzer. 
The whole system burns at most 2.3\,W but 
the EdgeDRNN only consumes 66\,mW. 
It is interesting to note that the
DRAM power is about 8X more than the RNN logic. 
This result clearly shows that the RNN computation is memory dominated.

%---------------------------------------------------------------------------------------------
\section{Comparison}
\label{sec:compare}
%---------------------------------------------------------------------------------------------
\subsection{Comparison with FPGA RNN Accelerators}
\begin{table*}[!t]
\caption{Comparison with state-of-the-art FPGA RNN accelerators.}
\label{tab:compare_fpga}
\centering
\begin{threeparttable}
\begin{tabular}{|l|c|c|c|c|c|}
\hline
\textbf{Platform}                                                                                          & \textbf{This Work} & \textbf{BBS}~\cite{bbs2019}    & \textbf{DeltaRNN}~\cite{deltarnn} & \textbf{ESE}~\cite{han2017ese} & \textbf{DeepRnn}~\cite{Chang2017}      \\ \hline
\textbf{FPGA}                                                                                              & XC7Z007S           & Arria 10 GX1150 & XC7Z100           & XCKU060      & XC7Z045               \\ \hline
\textbf{Dev. Kit Cost}                                                                                     & \$89               & \$4,495         & \$2,295           & \$3,295      & \$2,495               \\ \hline
\textbf{Weight Storage}                                                                                    & Off-chip           & On-chip \& off-chip         & On-chip           & Off-chip     & On-chip \& off-chip               \\ \hline
\textbf{\begin{tabular}[c]{@{}l@{}}Bit Precision\\ (Activation/Weight/Index)\end{tabular}}                 & INT 16/8/0           & INT 16/16/4       & INT 16/16/0         & INT 16/12/4    & INT 16/16/0             \\ \hline
\textbf{Sparsity Type}                                                                                     & Temporal           & Weight          & Temporal          & Weight       & -                  \\ \hline
\textbf{\tnote{1}~Effective Sparsity}                                                                                & 90.0\%             & 87.5\%          & 88.2\%            & 88.7\%       & -                     \\ \hline
\textbf{Frequency (MHz)}                                                                                   & 125                & 200             & 125               & 200          & 142                   \\ \hline
\textbf{\begin{tabular}[c]{@{}l@{}}DRAM Interface Bit Width\\ for Weight Fetch\end{tabular}}                                                                          & 64                 & -               & -                 & 512          & -                   \\ \hline
\textbf{Number of MACs (Batch-1)}                                                                          & 8                  & 4096            & 768               & 32           & 256                   \\ \hline
\textbf{\tnote{2}~Peak Throughput (GOp/s)}                                                                           & 2                  & \textbf{1638.4}          & 192               & 12.8         & 4.5                   \\ \hline
\textbf{\begin{tabular}[c]{@{}l@{}}Effective Batch-1\\ Throughput (GOp/s)\end{tabular}}                    & 20.2               & \textbf{2432.8}          & 1198.3            & 78.8         & 0.7                   \\ \hline
\textbf{\tnote{3}~MAC Utilization}                                                                                    & \textbf{1010}\%             & 150\%           & 630\%             & 620\%        & 15\%                  \\ \hline
\textbf{\begin{tabular}[c]{@{}l@{}}\tnote{4}~Memory-bounded\\ Peak Throughput (GOp/s)\end{tabular}}                                                                           & \textbf{2.0}                  & 1.3          & 2.0               & 1.3         & 2.0                   \\ \hline
\textbf{\begin{tabular}[c]{@{}l@{}}\tnote{5}~Normalized Effective Batch-1\\ Throughput (GOp/s)\end{tabular}}         & \textbf{20.2}               & $\leq$ 10.7     & $\leq$ 17.0       & $\leq$ 11.5  & $\leq$ 2              \\ \hline
\textbf{Wall Plug Power (W)}                                                                               & \textbf{2.3}                & 19.1            & 7.3               & 41.0+PC     & -                     \\ \hline
\textbf{\begin{tabular}[c]{@{}l@{}}Batch-1 Wall Plug\\ Power Efficiency (GOp/s/W)\end{tabular}}            & 8.8                & 127.4           & \textbf{164.2}             & 1.9          & -                     \\ \hline
\textbf{\begin{tabular}[c]{@{}l@{}}\tnote{5}~Normalized Batch-1 Wall Plug\\ Power Efficiency (GOp/s/W)\end{tabular}} & \textbf{8.8}                & $\leq$ 4.7      & $\leq$ 7.4        & $\leq$ 5.0   & - \\ \hline
\end{tabular}
\begin{tablenotes}\footnotesize
\item[1] The effective sparsity of EdgeDRNN \& DeltaRNN is calculated by Eq.~\ref{eq:4}.
\item[2] Peak throughput is calculated by Eq.~\ref{eq:6}.
\item[3] MAC utilization is the ratio between batch-1 throughput and the peak throughput of the accelerator.
\item[4] Memory-bounded peak throughput is calculated by Eq.~\ref{eq:9}.
\item[5] Normalized to the same frequency, DRAM interface bit width for weight fetch, number of MACs and activation \& weight bit precision with EdgeDRNN. We assume the normalized numbers are obtained on the same MiniZed board and assume they have the same wall plug power consumption with EdgeDRNN. Detailed discussion is in Section V.A.

% \item[3] Peak DRAM BW computed from bus width times memory frequency.
\end{tablenotes}
\end{threeparttable}
\end{table*}

Table~\ref{tab:compare_fpga} compares EdgeDRNN with other state-of-the-art FPGA RNN accelerators.
Both BBS~\cite{bbs2019} and DeltaRNN were optimized for batch-1 inference by using all MACs for a single input sample. BBS can use DRAM to support large networks and has the highest batch-1 throughput among all accelerators; however the reported throughput number was obtained by buffering the whole network by using expensive on-chip memory. 
After compression, the network has around 0.8\,MB parameters, 
which can be buffered on large FPGAs like the GX1150 used by BBS, 
but it is still too expensive for edge hardware platforms (e.g. MiniZed has only 0.2\,MB on-chip memory). 
ESE~\cite{han2017ese} reuses weights fetched from off-chip memory to feed 1024 MACs for batch inference and achieved 2520\,GOp/s total throughput; however only 32 out of 1024 MACs were used for each input sample limiting its batch-1 throughput. 
Except for EdgeDRNN and DeepRnn~\cite{Chang2017}, other platforms are not designed for edge applications. 
BBS, DeltaRNN and ESE provide much higher throughput but their power consumption is around 3X-18X larger than EdgeDRNN and they require expensive FPGA development systems that are not very portable.  
By contrast,
the small number of processing elements in EdgeDRNN is intentionally chosen to match the available memory bandwidth of the DRAM interface, 
since there is no point in having idle PEs.

To fairly compare architectures without the influence of different specifications of FPGA platforms, 
it makes sense to normalize the batch-1 throughput and other corresponding numbers of accelerators to the same number of PEs ($K=8$)\footnote{Each PE has a single MAC unit.}, 
clock frequency ($f_{pl}=125$\,MHz), DRAM interface bit width for weight fetch (64-bit) and bit precision of weights (INT8) \& activations (INT16) as used by EdgeDRNN.
We also assume that the normalized platforms are implemented on MiniZed having the same power consumption of EdgeDRNN. 
The normalized batch-1 throughput $\nu_{\rm Eff, Norm}$ is defined below:
\begin{equation}
\begin{aligned}
    \nu_{\rm Peak, Mem}&=2\cdot f_{pl}\cdot\frac{W_{\rm DRAM}}{W_{\rm Weight}+W_{\rm Index}} \\
    \nu_{\rm Eff, Norm}&=\nu_{\rm Peak, Mem}\cdot\frac{1}{1-\Gamma_{\rm Eff}}
\end{aligned}
\label{eq:9}
\end{equation}
where $\nu_{\rm Peak, Mem}$ is the memory-bounded peak throughput and $W_{\rm Index}$ is the bit width of the nonzero element index. 
To exploit weight sparsity by skipping zero elements in the weights, indices of nonzero weight elements have to be used and introduces off-chip memory overhead. 
Both BBS and ESE use $W_{\rm Index}=4$ for their tested networks. 
EdgeDRNN and DeltaRNN only need indices of valid columns corresponding to nonzero delta state vector elements, and they are calculated on-chip without introducing off-chip memory overhead; thus, $W_{\rm Index}=0$ for EdgeDRNN and DeltaRNN. 
In this normalization process, we assume the ideal case, in which normalized platforms reach the memory-bounded peak throughput and can fully utilize sparsity. Thus, Eq.~\ref{eq:9} gives the upper bound throughput value of the normalized platform.

Table~\ref{tab:compare_fpga} shows that EdgeDRNN achieves the highest normalized throughput, and an even higher normalized throughput than our previous BRAM-based DeltaRNN because of the improved pipeline and higher sparsity achieved. 
Compared with BBS, EdgeDRNN achieves only a small fraction of the total batch-1 throughput, but the normalization makes it clear that BBS achieves its high throughput by using on-chip BRAM, a huge number of MACs, and a higher clock frequency.
Among all the accelerators, EdgeDRNN also shows the highest effective MAC utilization and the lowest wall plug power. 
Finally, the EdgeDRNN FPGA development kit is a factor of at least 25X cheaper than other FPGA RNNs, and the cost is comparable to the cheapest edge AI accelerators.
%------------------------------------------------------------------
\subsection{Architectural Comparison}
\begin{figure}[!t]
	\centering
	\includegraphics[width=0.48\textwidth]{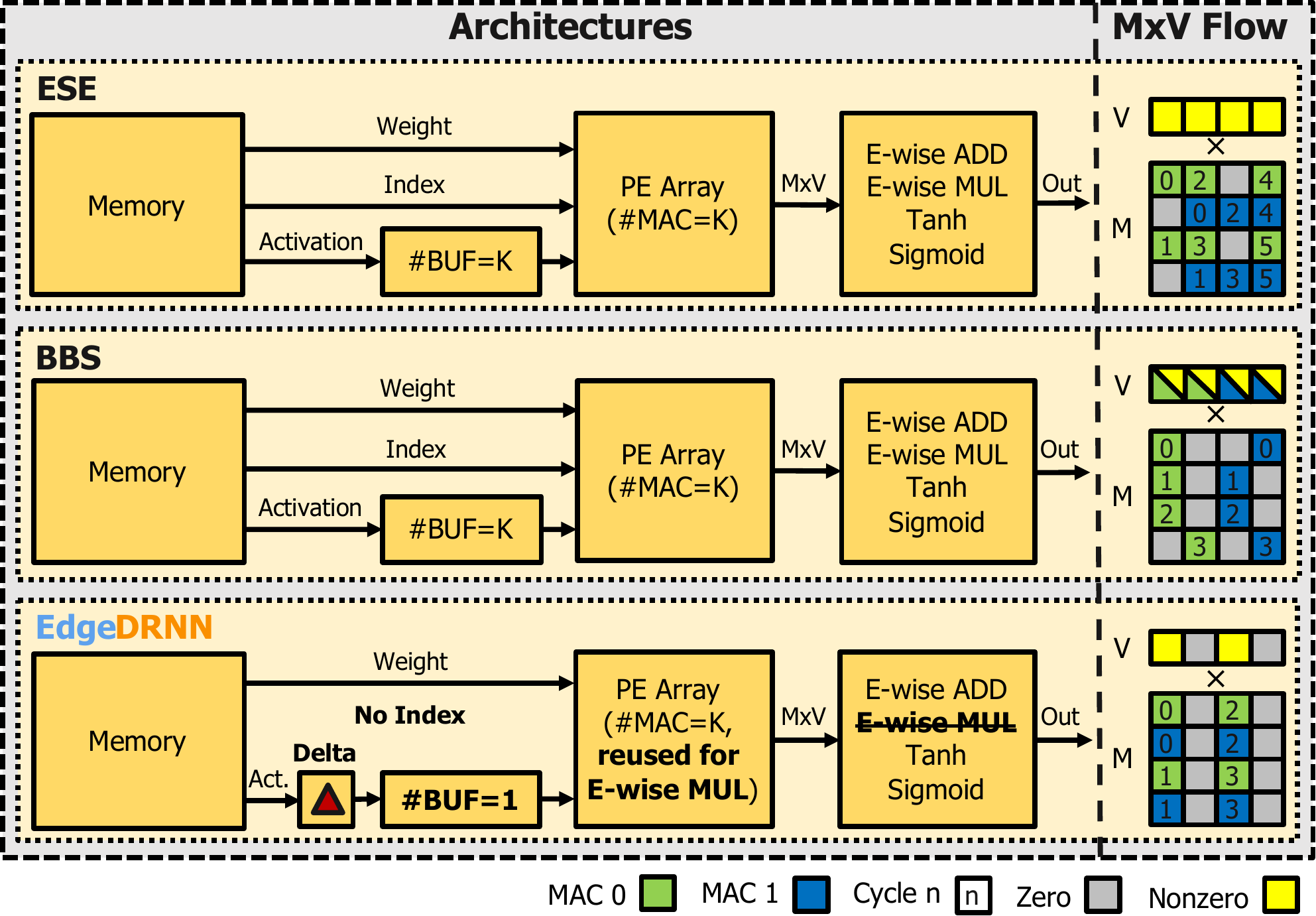}
	\caption{Architecture comparison and MxV Flow of ESE, BBS and EdgeDRNN. Bold text indicates differences. Colors in the MxV Flow part indicate MAC unit, and zero and nonzero activation values. The text indicates the clock cycle.}
	\label{fig:archicturecomparision}
\end{figure}

Fig.~\ref{fig:archicturecomparision} compares the architecture and MxV flow of EdgeDRNN with BBS and ESE.
We compare EdgeDRNN with ESE and BBS because they are also FPGA RNN accelerators using DRAM with high reported throughput.
Both ESE and BBS exploit weight sparsity with load balancing techniques.
\subsubsection{ESE}
In ESE, interleaving rows of matrix elements are assigned to MAC units in the PE array and the MxV is computed column by column. 
To balance the workload and exploit weight sparsity better, the network weight matrix is pruned so that the number of nonzero elements assigned to each MAC unit is the same for the whole matrix. 
$K$ activation buffers (\textbf{BUF}) are required for $K$ MAC units, which immediately execute operations when nonzero activations and weights are available.
\subsubsection{BBS}
BBS balances the workload using structured pruning. 
Rows of a weight matrix are split into banks of equal length. 
Their pruning method forces the numbers of nonzero values to be the same across all banks. By assigning the same number of row banks to each MAC unit in the PE array, the workload is balanced. As shown on the right side of Fig.~\ref{fig:archicturecomparision}, each row of the matrix is equally divided into two banks respectively for MAC\,0 and MAC\,1 and the computation is done row by row.
In this case, each MAC receives different activation elements and $K$ BUFs are required for $K$ MAC units.
BBS also supports the buffering of a partial weight matrix on-chip to enhance throughput, which is useful for large FPGA platforms.
The reported batch-1 throughput of BBS in Table~\ref{tab:compare_fpga} is obtained with all network parameters on-chip, which is not practical on a small FPGA platform like MiniZed that has only 0.2\,MB on-chip memory.

\subsubsection{EdgeDRNN}
%---------------------------------------------------------------------------------------------------------------
\begin{table*}[!t]
\caption{Comparison with commercial edge AI platforms and a desktop GTX 1080 GPU as benchmarked on the spoken digit recognition task.}
\label{tab:compare_asic}
\centering

\resizebox{1.00\textwidth}{!}
{
\begin{threeparttable}
\begin{tabular}{|l|c|c|c|c|c|c|c|c|c|c|}
\hline
\textbf{Platform}                                                                               & \multicolumn{3}{c|}{\textbf{This Work}}                                  & \textbf{NCS2~\cite{ncs2}}          & \multicolumn{2}{c|}{\textbf{Jetson Nano~\cite{jetson_nano}}} & \multicolumn{2}{c|}{\textbf{Jetson TX2~\cite{jetson_tx2}}} & \multicolumn{2}{c|}{\textbf{GTX 1080~\cite{gtx1080}}} \\ \hline
\textbf{Chip}                                                                                   & \multicolumn{3}{c|}{XC7Z007S}                                            & Myriad X               & \multicolumn{2}{c|}{Tegra X1}             & \multicolumn{2}{c|}{Tegra X2}            & \multicolumn{2}{c|}{GP104}             \\ \hline
\textbf{Dev. Kit Cost}                                                                          & \multicolumn{3}{c|}{\$89}                                                & \$69                   & \multicolumn{2}{c|}{\$99}                 & \multicolumn{2}{c|}{\$411}               & \multicolumn{2}{c|}{\$500+PC}          \\ \hline
\textbf{DRAM Type (Bus Width)}                                                                  & \multicolumn{3}{c|}{DDR3L (16-bit)}                                      & -                      & \multicolumn{2}{c|}{LPDDR4 (64-bit)}      & \multicolumn{2}{c|}{LPDDR4 (128-bit)}    & \multicolumn{2}{c|}{GDDR5X (256-bit)}  \\ \hline
\textbf{DRAM Bandwidth (GB/s)}                                                                  & \multicolumn{3}{c|}{1.0}                                                 & -                      & \multicolumn{2}{c|}{25.6}                 & \multicolumn{2}{c|}{59.7}                & \multicolumn{2}{c|}{320}               \\ \hline
\textbf{Test Network}                                                                           & \multicolumn{3}{c|}{2L-768H-DeltaGRU}                                    & 2L-664H-LSTM           & \multicolumn{6}{c|}{2L-768H-GRU}                                                                                              \\ \hline
\textbf{\#Parameters}                                                                           & \multicolumn{3}{c|}{5.4 M}                                               & 5.4 M                  & \multicolumn{6}{c|}{5.4 M}                                                                                                    \\ \hline
\textbf{Bit Precision (A/W)}                                                                    & \multicolumn{3}{c|}{INT16/8}                                             & FP16                   & FP32                & FP16                & FP32                & FP16               & FP32               & FP16              \\ \hline
\multirow{2}{*}{\textbf{WER on TIDIGITS}}                                                       & $\Theta=0\mathbf{x}00$ & $\Theta=0\mathbf{x}08$ & $\Theta=0\mathbf{x}40$ & \multirow{2}{*}{1.1\%} & \multicolumn{6}{c|}{\multirow{2}{*}{0.8\%}}                                                                                   \\ \cline{2-4}
                                                                                                & 0.7\%                  & 0.8\%                  & 1.3\%                  &                        & \multicolumn{6}{c|}{}                                                                                                         \\ \hline
\textbf{Latency ($\mu$s)}                                                                       & 2633                   & 1673                   & \textbf{536}           & 3,588                  & 5,757               & 4,356               & 3,124               & 2,693              & \textbf{527}       & \textbf{484}      \\ \hline
\textbf{\begin{tabular}[c]{@{}l@{}}Effective Batch-1\\ Throughput (GOp/s)\end{tabular}}         & 4.1                    & 6.5                    & \textbf{20.2}          & 3.0                    & 1.9                 & 2.5                 & 3.5                 & 4.0                & \textbf{20.5}      & \textbf{22.3}     \\ \hline
\textbf{\tnote{1}~Wall Plug Power (W)}                                                                    & \multicolumn{3}{c|}{\textbf{2.3}}                                                 & \textbf{1.7}                    & 7.2                 & 7.1                 & 8.2                 & 8.1                & 96.6+PC            & 82.2+PC           \\ \hline
\textbf{\begin{tabular}[c]{@{}l@{}}Batch-1 Wall Plug\\ Power Efficiency (GOp/s/W)\end{tabular}} & 1.8                    & 2.8                    & \textbf{8.8}           & 1.8                    & 0.3                 & 0.4                 & 0.4                 & 0.5                & 0.2                & 0.3               \\ \hline
\end{tabular}
\begin{tablenotes}\footnotesize
\item[1] EdgeDRNN power was measured by a USB power meter. Power numbers of Jetson Nano and Jetson TX2 boards are measured by a Voltcraft 4500ADVANCED Energy Monitor. Power of GTX 1080 was measured by the \textit{nvidia-smi} utility.

% \item[3] Peak DRAM BW computed from bus width times memory frequency.
\end{tablenotes}
\end{threeparttable}
}
\end{table*}
%---------------------------------------------------------------------------------------------------------------
Unlike ESE and BBS, EdgeDRNN includes an extra unit to compute delta state vectors.
Similar to ESE, EdgeDRNN also assigns interleaving rows to MAC units and computes MxV column by column; however, all MAC units share the same delta state vector elements; thus, only 1 BUF (D-FIFO) is required. 

Both ESE and BBS require indices of nonzero weight elements to realize zero-skipping. The indices cause overhead on memory access, reducing effective memory bandwidth. 
EdgeDRNN skips whole columns of computation and indices of valid columns are calculated on-the-fly to avoid memory overhead.

Moreover, ESE and BBS require extra Element-wise (\textbf{E-wise}) multiplication units for the RNN activation $h_{t}$ generation after MxV. EdgeDRNN reuses multipliers in the PE array by time-division multiplexing to save DSP and LUT resources. Element-wise addition is done by reusing adders in the PE array and also using a single 16-bit adder per PE, as shown in Fig.~\ref{fig:pe}.

Our previous work, DeltaRNN~\cite{deltarnn}, achieved high batch-1 throughput and MAC utilization with temporal sparsity, but it stored all network parameters on chip, making it unscalable. Meanwhile, EdgeDRNN is designed to match the external memory bandwidth available on any FPGA platform with external DRAM. 
The small number of MAC units along tall concatenated weight matrix columns, as shown in Fig.~\ref{fig:mxv}, makes the burst length long enough to maintain high DRAM controller efficiency for large networks.
\begin{figure}[!t]
	\centering
	\includegraphics[width=0.48\textwidth]{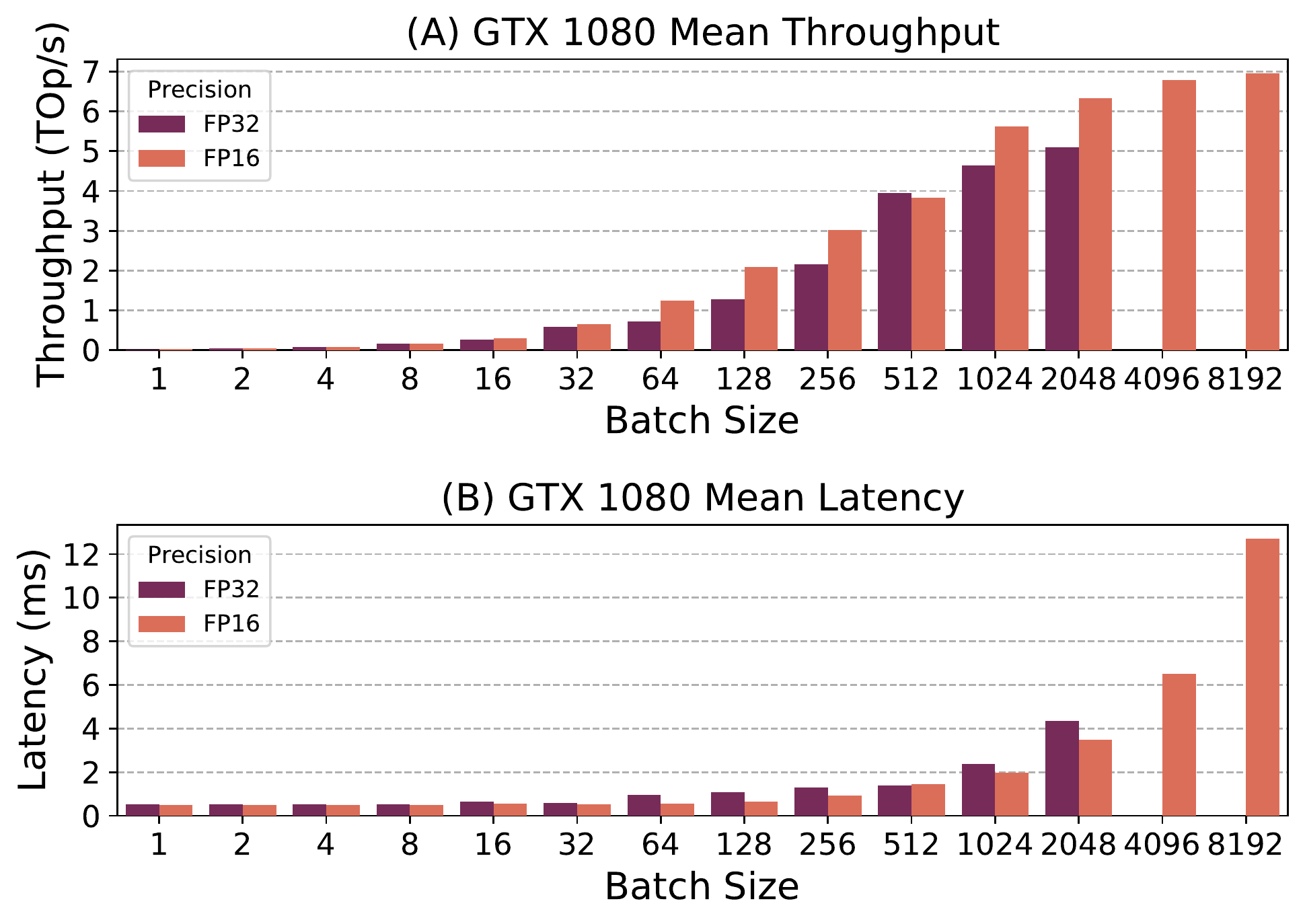}
	\caption{Measured mean throughput (A)  \& latency (B) of GTX 1080 versus different batch sizes up to the maximum batch size that fits into the available DRAM.}
	\label{fig:gpu_bench}
\end{figure}

\subsection{Comparison with an SNN Processor}
We compare the performance metrics of TrueNorth~\cite{truenorth}, an application-specific integrated circuit (\textbf{ASIC}) SNN processor, on the TIDIGITS dataset. 
The system is reported to dissipate 38.6\,mW power by using a feature extraction method that can be implemented directly on TrueNorth~\cite{truenorth_speech}. We cannot easily compare the power numbers of this ASIC processor  with the power dissipated by an FPGA which is a more general-purpose platform. 
To run TrueNorth, an interfacing FPGA that burns several Watts is needed so the system power would much higher. 
The reported accuracy from their work is only 95\% which is lower than the 99\% accuracy achieved by the quantized delta network reported in our previous work~\cite{neil2016delta}.

\subsection{Comparison with Commercial Platforms}
\begin{figure}[!t]
	\centering
	\includegraphics[width=0.48\textwidth]{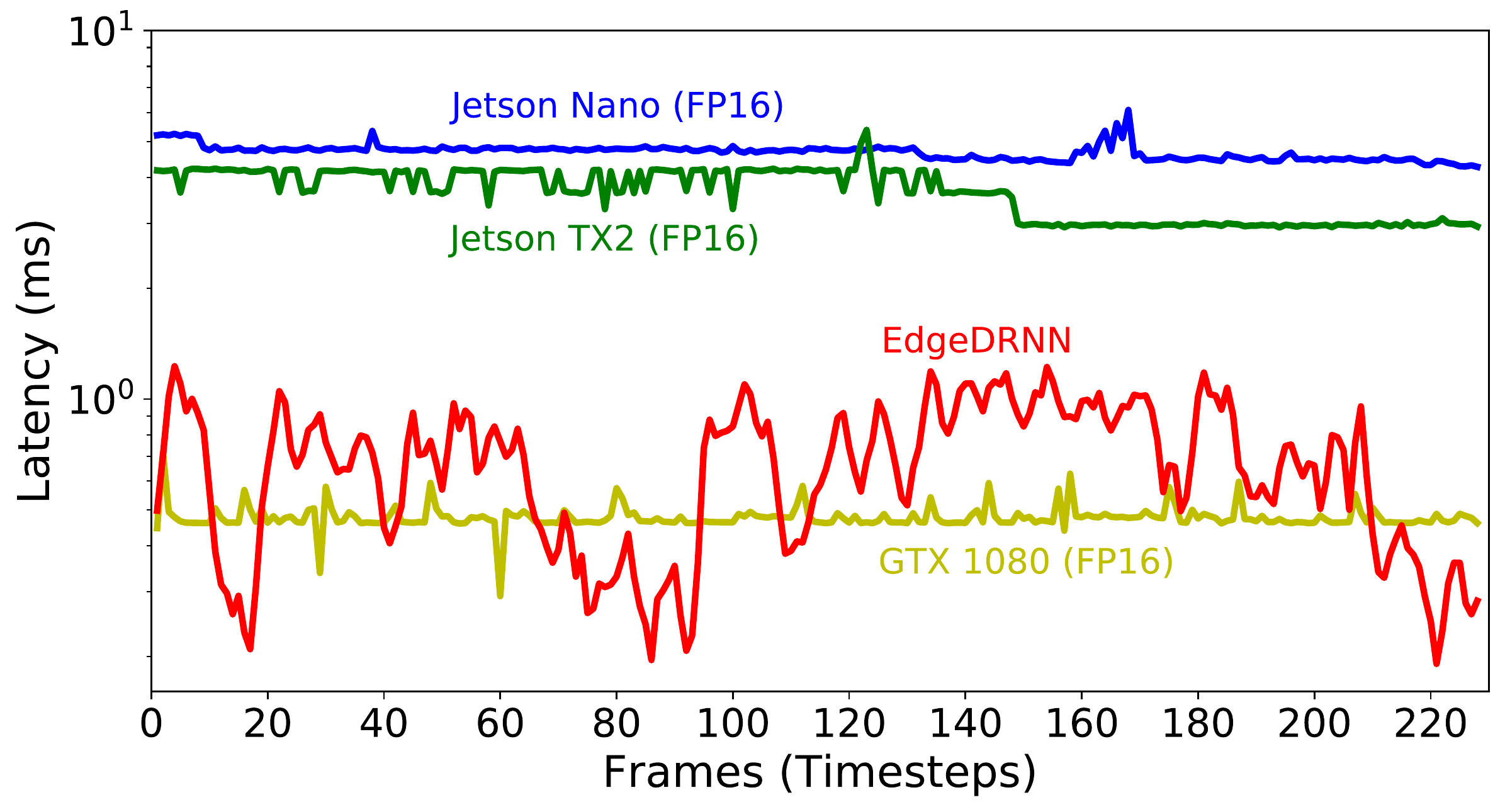}
	\caption{Measured batch-1 latency per frame of a sample (25896O4A.WAV) from the \textsl{TIDIGITS} test set benchmarked on EdgeDRNN (INT16 \& INT8) and GPUs (FP16) with the first 50 timesteps excluded.}
	\label{fig:latency}
\end{figure}

Table~\ref{tab:compare_asic} compares EdgeDRNN with popular commercial platforms, including their cost and memory system bandwidth. 
All platforms are benchmarked on the same spoken digit recognition task (first 10,000 timesteps of the \textit{TIDIGITS} test set) using networks of the same size, except that the Intel Compute Stick 2 (NCS2) 
does not support GRU and was benchmarked with an LSTM 
network with a similar parameter count and trained with the same hyperparameters.
The latency requirement of the recognition task is 10\,ms which is determined by the frame stride. 
To meet this requirement, frames cannot be concatenated into a single tensor. 
The computation of the RNN is executed when there is a new frame. 

For benchmark of GPUs, we used GRUs because we found that latency numbers of both FP32 and FP16 cuDNN GRU implementations are 3X lower than that of running the DeltaGRU algorithm using the NVIDIA cuSPARSE library. 
In addition, we removed peripheral devices from the Jetson board with the exclusion of the needed Ethernet cable to the PC. 
Because GPUs also need time to boost their clock frequency and to allocate memory, the first 50 timesteps of the test sequence are excluded.  
The power efficiency results show that EdgeDRNN still achieves over 5X higher system power efficiency compared to commercial ASIC and GPU products.

GPUs are throughput-oriented architectures suitable for neural network training with large batch sizes; 
however, 
it is not optimal for edge inference where batch-1 throughput is critical for achieving low latency. 
The claimed peak FP32 throughput of Jetson Nano~\cite{jetson_nano}, Jetson TX2~\cite{jetson_tx2} and GTX 1080~\cite{gtx1080} are respectively 0.5\,TOp/s, 0.8\,TOp/s and 9\,TOp/s while the measured batch-1 throughput are only 1.9\,GOp/s, 3.5\,GOp/s and 20.5\,GOp/s. 
The low batch-1 throughput of GPUs is because weights fetched from off-chip DRAM cannot be reused to fully utilize GPU cores. 
Fig.~\ref{fig:gpu_bench}A shows the throughput of GTX 1080 approaches the claimed peak throughput with large batch sizes due to more weight data reuse;
however, increasing batch size also causes worse latency numbers as shown in Fig.~\ref{fig:gpu_bench}B.
FP16 outperforms FP32 because of the smaller memory bottleneck.

Fig.~\ref{fig:latency}  compares latency per frame on a test set sample. EdgeDRNN latency is lower during the silent or quieter periods (e.g. between 60\,s and 80\,s) when the input is changing slowly.
EdgeDRNN is as quick as the desktop 1080 GPU and 5X quicker than the other platforms, despite having a DRAM bandwidth that is orders of magnitude slower. 
\begin{figure}[!t]
	\centering
	\includegraphics[width=0.48\textwidth]{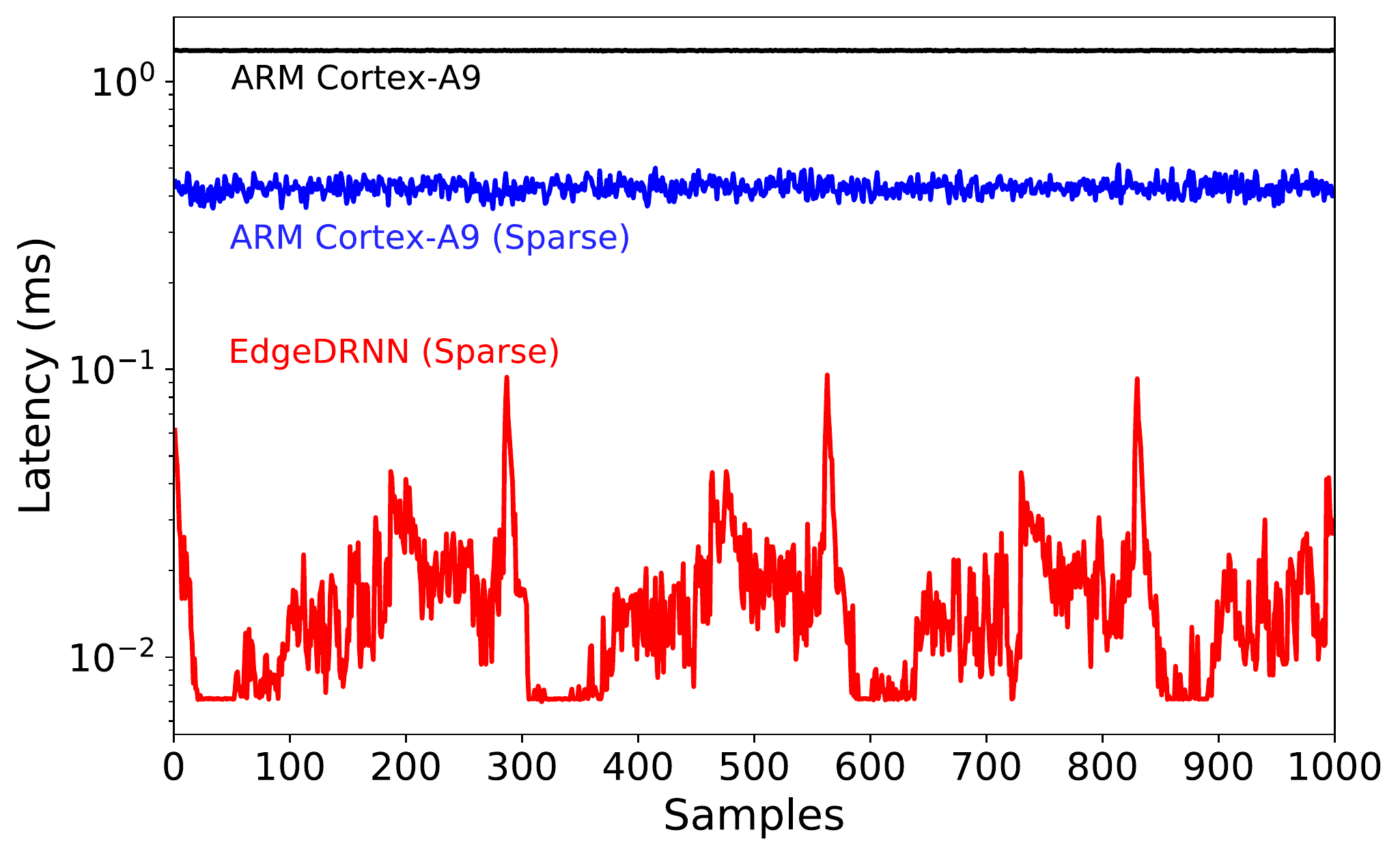}
	\caption{Measured hardware latency per sample on EdgeDRNN and embedded processor using the 2L-128H-DeltaGRU network used in the real-time control demonstration of AMPRO~\cite{ampro}.}
	\label{fig:latency_ampro}
\end{figure}

In~\cite{ampro}, we reported that EdgeDRNN ran the RNN for robotic control about 51X faster than the embedded BeagleBone Black platform with a ARM Cortex-A8 CPU, while burning about the same total power of 2W. 
Moreover, to compare the performance of EdgeDRNN and the ARM Cortex-A9 CPU on the PS side of MiniZed, we took the same 2L-128H-DeltaGRU network used in our previous real-time control demonstration~\cite{ampro} and measured the latency per frame on 1 minute test data (sample rate = 200 Hz), 
which are 1000 frames of motor encoder readings. 
Fig.~\ref{fig:latency_ampro} shows the latency of the ARM CPU and EdgeDRNN. 
The mean latency of the ARM CPU is 1281\,$\mu$s without sparsity and 428\,$\mu$s with sparsity. 
The mean latency of EdgeDRNN with sparsity is 16\,$\mu$s, therefore EdgeDRNN is 27X faster than the ARM CPU which exploits temporal sparsity in the same network. In the case of the robotic task, EdgeDRNN runs the network 300X faster than the required maximum latency of 5ms.

%*********************************************************************************************
\section{Conclusion}
\label{sec:conclude}
%*********************************************************************************************
The 2\,W EdgeDRNN runs batch-1 RNNs as fast as a 200\,W GPU+PC, 
and its power efficiency is at least a factor of 4X higher than any of the commercial edge AI platforms in the benchmark.
We found that the batch-1 RNN throughput numbers of commercial GPUs are a factor of over 100X less than their claimed peak throughput.
Using the delta network to exploit temporal sparsity allows a modest number of 8 PEs to achieve an effective 162\,Op per clock cycle, 
equivalent to an MAC utilization efficiency of over 1000\%.
EdgeDRNN uses a standard AXI4 interface for weight fetches; thus it can be scaled up to larger FPGA platforms by simply increasing the number of PEs to match the available memory bandwidth provided by on-chip BRAM or off-chip DRAM.
Thus using temporal sparsity in delta activation vectors allows the arithmetic units 
on this task to effectively compute 10X more operations with the same amount of memory access.

The delta threshold $\Theta$ allows instantaneous trade-off of accuracy versus latency. 
Future work could exploit a dynamic trade-off of accuracy versus latency to quickly converge onto optimal values in a guided search.

\ifCLASSOPTIONcaptionsoff
  \newpage
\fi
%*********************************************************************************************
% \bibliographystyle{IEEEtran}
% \bibliography{IEEEabrv}

\begin{thebibliography}{10}
\providecommand{\url}[1]{#1}
\csname url@samestyle\endcsname
\providecommand{\newblock}{\relax}
\providecommand{\bibinfo}[2]{#2}
\providecommand{\BIBentrySTDinterwordspacing}{\spaceskip=0pt\relax}
\providecommand{\BIBentryALTinterwordstretchfactor}{4}
\providecommand{\BIBentryALTinterwordspacing}{\spaceskip=\fontdimen2\font plus
\BIBentryALTinterwordstretchfactor\fontdimen3\font minus
  \fontdimen4\font\relax}
\providecommand{\BIBforeignlanguage}[2]{{%
\expandafter\ifx\csname l@#1\endcsname\relax
\typeout{** WARNING: IEEEtran.bst: No hyphenation pattern has been}%
\typeout{** loaded for the language `#1'. Using the pattern for}%
\typeout{** the default language instead.}%
\else
\language=\csname l@#1\endcsname
\fi
#2}}
\providecommand{\BIBdecl}{\relax}
\BIBdecl

\bibitem{graves2012}
A.~Graves, ``Sequence transduction with recurrent neural networks,'' \emph{ICML Representation Learning Workshop}, 2012.

\bibitem{Graves2013}
A.~{Graves}, A.~{Mohamed}, and G.~{Hinton}, ``Speech recognition with deep
  recurrent neural networks,'' in \emph{2013 IEEE International Conference on
  Acoustics, Speech and Signal Processing}, 2013, pp. 6645--6649.

\bibitem{funahashi_approximation_1993}
\BIBentryALTinterwordspacing
K.-i. Funahashi and Y.~Nakamura, ``Approximation of dynamical systems by
  continuous time recurrent neural networks,'' \emph{Neural Networks}, vol.~6,
  no.~6, pp. 801 -- 806, 1993. [Online]. Available:
  \url{http://www.sciencedirect.com/science/article/pii/S089360800580125X}
\BIBentrySTDinterwordspacing

\bibitem{chow_recurrent_1998}
T.~Chow and Y.~Fang, ``A recurrent neural-network-based real-time learning
  control strategy applying to nonlinear systems with unknown dynamics,''
  \emph{IEEE Transactions on Industrial Electronics}, vol.~45, no.~1, pp.
  151--161, Feb. 1998, conference Name: IEEE Transactions on Industrial
  Electronics.

\bibitem{lstm_hoch97}
\BIBentryALTinterwordspacing
S.~Hochreiter and J.~Schmidhuber, ``Long short-term memory,'' \emph{Neural
  Comput.}, vol.~9, no.~8, pp. 1735--1780, Nov. 1997. [Online]. Available:
  \url{http://dx.doi.org/10.1162/neco.1997.9.8.1735}
\BIBentrySTDinterwordspacing

\bibitem{gru_og}
\BIBentryALTinterwordspacing
K.~Cho, B.~van Merri{\"{e}}nboer, {\c C}.~G{\"{u}}l{\c c}ehre, D.~Bahdanau,
  F.~Bougares, H.~Schwenk, and Y.~Bengio, ``Learning phrase representations
  using {RNN} encoder--decoder for statistical machine translation,'' in
  \emph{Proceedings of the 2014 Conference on Empirical Methods in Natural
  Language Processing (EMNLP)}, Oct. 2014, pp. 1724--1734. [Online]. Available:
  \url{http://www.aclweb.org/anthology/D14-1179}
\BIBentrySTDinterwordspacing

\bibitem{ampro}
C.~{Gao}, R.~{Gehlhar}, A.~D. {Ames}, S.~C. {Liu}, and T.~{Delbruck},
  ``Recurrent neural network control of a hybrid dynamical transfemoral
  prosthesis with edgedrnn accelerator,'' in \emph{2020 IEEE International
  Conference on Robotics and Automation (ICRA)}, 2020, pp. 5460--5466.

\bibitem{chen2019}
J.~Chen and X.~Ran, ``Deep {Learning} {With} {Edge} {Computing}: {A}
  {Review},'' \emph{Proceedings of the IEEE}, vol. 107, no.~8, pp. 1655--1674,
  Aug. 2019.

\bibitem{han2017ese}
S.~Han, J.~Kang, H.~Mao, Y.~Hu, X.~Li, Y.~Li, D.~Xie, H.~Luo, S.~Yao, Y.~Wang
  \emph{et~al.}, ``{ESE: E}fficient speech recognition engine with sparse
  {LSTM} on {FPGA},'' in \emph{Proceedings of the 2017 ACM/SIGDA International
  Symposium on Field-Programmable Gate Arrays}.\hskip 1em plus 0.5em minus
  0.4em\relax ACM, 2017, pp. 75--84.

\bibitem{bbs2019}
\BIBentryALTinterwordspacing
S.~Cao, C.~Zhang, Z.~Yao, W.~Xiao, L.~Nie, D.~Zhan, Y.~Liu, M.~Wu, and
  L.~Zhang, ``Efficient and effective sparse lstm on fpga with bank-balanced
  sparsity,'' in \emph{Proceedings of the 2019 ACM/SIGDA International
  Symposium on Field-Programmable Gate Arrays}, ser. FPGA ’19.\hskip 1em plus
  0.5em minus 0.4em\relax New York, NY, USA: Association for Computing
  Machinery, 2019, p. 63–72. [Online]. Available:
  \url{https://doi.org/10.1145/3289602.3293898}
\BIBentrySTDinterwordspacing

\bibitem{deltarnn}
\BIBentryALTinterwordspacing
C.~Gao, D.~Neil, E.~Ceolini, S.-C. Liu, and T.~Delbruck, ``{DeltaRNN}: A
  power-efficient recurrent neural network accelerator,'' in \emph{Proceedings
  of the 2018 ACM/SIGDA International Symposium on Field-Programmable Gate
  Arrays}, ser. FPGA ’18.\hskip 1em plus 0.5em minus 0.4em\relax New York,
  NY, USA: Association for Computing Machinery, 2018, p. 21–30. [Online].
  Available: \url{https://doi.org/10.1145/3174243.3174261}
\BIBentrySTDinterwordspacing

\bibitem{neil2016delta}
\BIBentryALTinterwordspacing
D.~Neil, J.~Lee, T.~Delbr{\"{u}}ck, and S.-C. Liu, ``Delta networks for
  optimized recurrent network computation,'' in \emph{Proceedings of the 34th
  International Conference on Machine Learning, {ICML} 2017, Sydney, NSW,
  Australia, 6-11 August 2017}, 2017, pp. 2584--2593. [Online]. Available:
  \url{http://proceedings.mlr.press/v70/neil17a.html}
\BIBentrySTDinterwordspacing

\bibitem{gaoAICAS2020}
C.~{Gao}, A.~{Rios-Navarro}, X.~{Chen}, T.~{Delbruck}, and S.-C. {Liu},
  ``{EdgeDRNN}: Enabling low-latency recurrent neural network edge inference,''
  in \emph{2020 2nd IEEE International Conference on Artificial Intelligence
  Circuits and Systems (AICAS)}, 2020, pp. 41--45.

\bibitem{Chang2017}
A.~X.~M. Chang and E.~Culurciello, ``Hardware accelerators for recurrent neural
  networks on {FPGA},'' in \emph{2017 IEEE International Symposium on Circuits
  and Systems (ISCAS)}, May 2017, pp. 1--4.

\bibitem{Lee2016}
M.~{Lee}, K.~{Hwang}, J.~{Park}, S.~{Choi}, S.~{Shin}, and W.~{Sung},
  ``Fpga-based low-power speech recognition with recurrent neural networks,''
  in \emph{2016 IEEE International Workshop on Signal Processing Systems
  (SiPS)}, 2016, pp. 230--235.

\bibitem{Guan2017FCCM}
Y.~Guan, H.~Liang, N.~Xu, W.~Wang, S.~Shi, X.~Chen, G.~Sun, W.~Zhang, and
  J.~Cong, ``{FP-DNN: A}n automated framework for mapping deep neural networks
  onto {FPGAs with RTL-HLS} hybrid templates,'' in \emph{2017 IEEE 25th Annual
  International Symposium on Field-Programmable Custom Computing Machines
  (FCCM)}, April 2017, pp. 152--159.

\bibitem{Shin2017}
D.~Shin, J.~Lee, J.~Lee, and H.~Yoo, ``14.2 {DNPU}: An 8.1{TOPS/W}
  reconfigurable {CNN-RNN} processor for general-purpose deep neural
  networks,'' in \emph{2017 IEEE International Solid-State Circuits Conference
  (ISSCC)}, Feb 2017, pp. 240--241.

\bibitem{yangbnn2018}
M.~Yang, C.~Yeh, Y.~Zhou, J.~P. Cerqueira, A.~A. Lazar, and M.~Seok, ``A
  1$\mu${W} voice activity detector using analog feature extraction and digital
  deep neural network,'' in \emph{2018 IEEE International Solid - State
  Circuits Conference - (ISSCC)}, Feb 2018, pp. 346--348.

\bibitem{stromatias2015robustness}
E.~Stromatias, D.~Neil, M.~Pfeiffer, F.~Galluppi, S.~B. Furber, and S.-C. Liu,
  ``Robustness of spiking deep belief networks to noise and reduced bit
  precision of neuro-inspired hardware platforms,'' \emph{Frontiers in
  Neuroscience}, vol.~9, p. 222, 2015.

\bibitem{Han2015}
\BIBentryALTinterwordspacing
S.~Han, H.~Mao, and W.~J. Dally, ``Deep compression: Compressing deep neural
  network with pruning, trained quantization and huffman coding,'' in \emph{4th
  International Conference on Learning Representations, {ICLR} 2016, San Juan,
  Puerto Rico, May 2-4, 2016, Conference Track Proceedings}, Y.~Bengio and
  Y.~LeCun, Eds., 2016. [Online]. Available:
  \url{http://arxiv.org/abs/1510.00149}
\BIBentrySTDinterwordspacing

\bibitem{Kadetotad2020}
D.~{Kadetotad}, S.~{Yin}, V.~{Berisha}, C.~{Chakrabarti}, and J.~{Seo}, ``An
  8.93 {TOPS/W} {LSTM} recurrent neural network accelerator featuring
  hierarchical coarse-grain sparsity for on-device speech recognition,''
  \emph{IEEE Journal of Solid-State Circuits}, pp. 1--1, 2020.

\bibitem{Judd2016}
P.~{Judd}, J.~{Albericio}, T.~{Hetherington}, T.~M. {Aamodt}, and
  A.~{Moshovos}, ``Stripes: Bit-serial deep neural network computing,'' in
  \emph{2016 49th Annual IEEE/ACM International Symposium on Microarchitecture
  (MICRO)}, 2016, pp. 1--12.

\bibitem{Bilaniuk2019}
O.~{Bilaniuk}, S.~{Wagner}, Y.~{Savaria}, and J.~{David}, ``Bit-slicing {FPGA}
  accelerator for quantized neural networks,'' in \emph{2019 IEEE International
  Symposium on Circuits and Systems (ISCAS)}, 2019, pp. 1--5.

\bibitem{Wang2017}
Z.~Wang, J.~Lin, and Z.~Wang, ``Accelerating recurrent neural networks: A
  memory-efficient approach,'' \emph{IEEE Transactions on Very Large Scale
  Integration (VLSI) Systems}, vol.~25, no.~10, pp. 2763--2775, Oct 2017.

\bibitem{Wang2018}
\BIBentryALTinterwordspacing
S.~Wang, Z.~Li, C.~Ding, B.~Yuan, Q.~Qiu, Y.~Wang, and Y.~Liang, ``{C-LSTM:
  E}nabling efficient {LSTM} using structured compression techniques on
  {FPGAs},'' in \emph{Proceedings of the 2018 ACM/SIGDA International Symposium
  on Field-Programmable Gate Arrays}, ser. FPGA '18.\hskip 1em plus 0.5em minus
  0.4em\relax New York, NY, USA: ACM, 2018, pp. 11--20. [Online]. Available:
  \url{http://doi.acm.org/10.1145/3174243.3174253}
\BIBentrySTDinterwordspacing

\bibitem{GaoISCAS2019}
C.~{Gao}, S.~{Braun}, I.~{Kiselev}, J.~{Anumula}, T.~{Delbruck}, and S.-C.
  {Liu}, ``Real-time speech recognition for {IoT} purpose using a delta
  recurrent neural network accelerator,'' in \emph{2019 IEEE International
  Symposium on Circuits and Systems (ISCAS)}, 2019, pp. 1--5.

\bibitem{leonard1993tidigits}
R.~G. Leonard and G.~Doddington, ``{TIDIGITS} speech corpus,'' \emph{Texas
  Instruments, Inc}, 1993.

\bibitem{burgues2018estimation}
J.~Burgu\'{e}s, J.~M. Jim\'{e}nez-Soto, and S.~Marco, ``Estimation of the limit
  of detection in semiconductor gas sensors through linearized calibration
  models,'' \emph{Analytica Chimica Acta}, vol. 1013, pp. 13 -- 25, 2 2018.

\bibitem{burgues2018multivariate}
J.~Burgu\'{e}s and S.~Marco, ``Multivariate estimation of the limit of
  detection by orthogonal partial least squares in temperature-modulated {MOX}
  sensors,'' \emph{Analytica Chimica Acta}, vol. 1019, pp. 49 -- 64, 2018.

\bibitem{sensorsgas}
S.~{Wang}, Y.~{Hu}, J.~{Burgués}, S.~{Marco}, and S.-C. {Liu}, ``Prediction of
  gas concentration using gated recurrent neural networks,'' in \emph{2020 2nd
  IEEE International Conference on Artificial Intelligence Circuits and Systems
  (AICAS)}, 2020, pp. 178--182.

\bibitem{axi_datamover}
\BIBentryALTinterwordspacing
Xilinx, ``{AXI} datamover.'' [Online]. Available:
  \url{https://www.xilinx.com/products/intellectual-property/axi_datamover.html}
\BIBentrySTDinterwordspacing

\bibitem{minized}
\BIBentryALTinterwordspacing
AVNET, ``{MiniZed}.'' [Online]. Available:
  \url{http://zedboard.org/product/minized}
\BIBentrySTDinterwordspacing

\bibitem{Graves2006}
\BIBentryALTinterwordspacing
A.~Graves, S.~Fern\'{a}ndez, F.~Gomez, and J.~Schmidhuber, ``Connectionist
  temporal classification: Labelling unsegmented sequence data with recurrent
  neural networks,'' in \emph{Proceedings of the 23rd International Conference
  on Machine Learning}, ser. ICML '06.\hskip 1em plus 0.5em minus 0.4em\relax
  New York, NY, USA: ACM, 2006, pp. 369--376. [Online]. Available:
  \url{http://doi.acm.org/10.1145/1143844.1143891}
\BIBentrySTDinterwordspacing

\bibitem{ncs2}
\BIBentryALTinterwordspacing
``Intel® neural compute stick 2 product specifications.'' [Online]. Available:
  \url{https://ark.intel.com/content/www/us/en/ark/products/140109/intel-neural-compute-stick-2.html}
\BIBentrySTDinterwordspacing

\bibitem{jetson_nano}
\BIBentryALTinterwordspacing
``\BIBforeignlanguage{en}{Jetson {Nano} {Developer} {Kit}},'' Mar. 2019.
  [Online]. Available:
  \url{https://developer.nvidia.com/embedded/jetson-nano-developer-kit}
\BIBentrySTDinterwordspacing

\bibitem{jetson_tx2}
\BIBentryALTinterwordspacing
``{Harness AI at the Edge with the Jetson TX2 Developer Kit},'' Aug 2019.
  [Online]. Available:
  \url{https://developer.nvidia.com/embedded/jetson-tx2-developer-kit}
\BIBentrySTDinterwordspacing

\bibitem{gtx1080}
\BIBentryALTinterwordspacing
``\BIBforeignlanguage{en-us}{{GeForce} {GTX} 1080 {Graphics} {Cards} from
  {NVIDIA} {GeForce}}.'' [Online]. Available:
  \url{https://www.nvidia.com/en-us/geforce/products/10series/geforce-gtx-1080/}
\BIBentrySTDinterwordspacing

\bibitem{truenorth}
F.~{Akopyan}, J.~{Sawada}, A.~{Cassidy}, R.~{Alvarez-Icaza}, J.~{Arthur},
  P.~{Merolla}, N.~{Imam}, Y.~{Nakamura}, P.~{Datta}, G.~{Nam}, B.~{Taba},
  M.~{Beakes}, B.~{Brezzo}, J.~B. {Kuang}, R.~{Manohar}, W.~P. {Risk},
  B.~{Jackson}, and D.~S. {Modha}, ``{TrueNorth: D}esign and tool flow of a 65
  {mW} 1 million neuron programmable neurosynaptic chip,'' \emph{IEEE
  Transactions on Computer-Aided Design of Integrated Circuits and Systems},
  vol.~34, no.~10, pp. 1537--1557, 2015.

\bibitem{truenorth_speech}
W.~{Tsai}, D.~R. {Barch}, A.~S. {Cassidy}, M.~V. {DeBole}, A.~{Andreopoulos},
  B.~L. {Jackson}, M.~D. {Flickner}, J.~V. {Arthur}, D.~S. {Modha},
  J.~{Sampson}, and V.~{Narayanan}, ``Always-on speech recognition using
  {TrueNorth}, a reconfigurable, neurosynaptic processor,'' \emph{IEEE
  Transactions on Computers}, vol.~66, no.~6, pp. 996--1007, 2017.

\end{thebibliography}

%*********************************************************************************************
\begin{IEEEbiography}[{\includegraphics[width=1in,height=1.25in,clip,keepaspectratio]{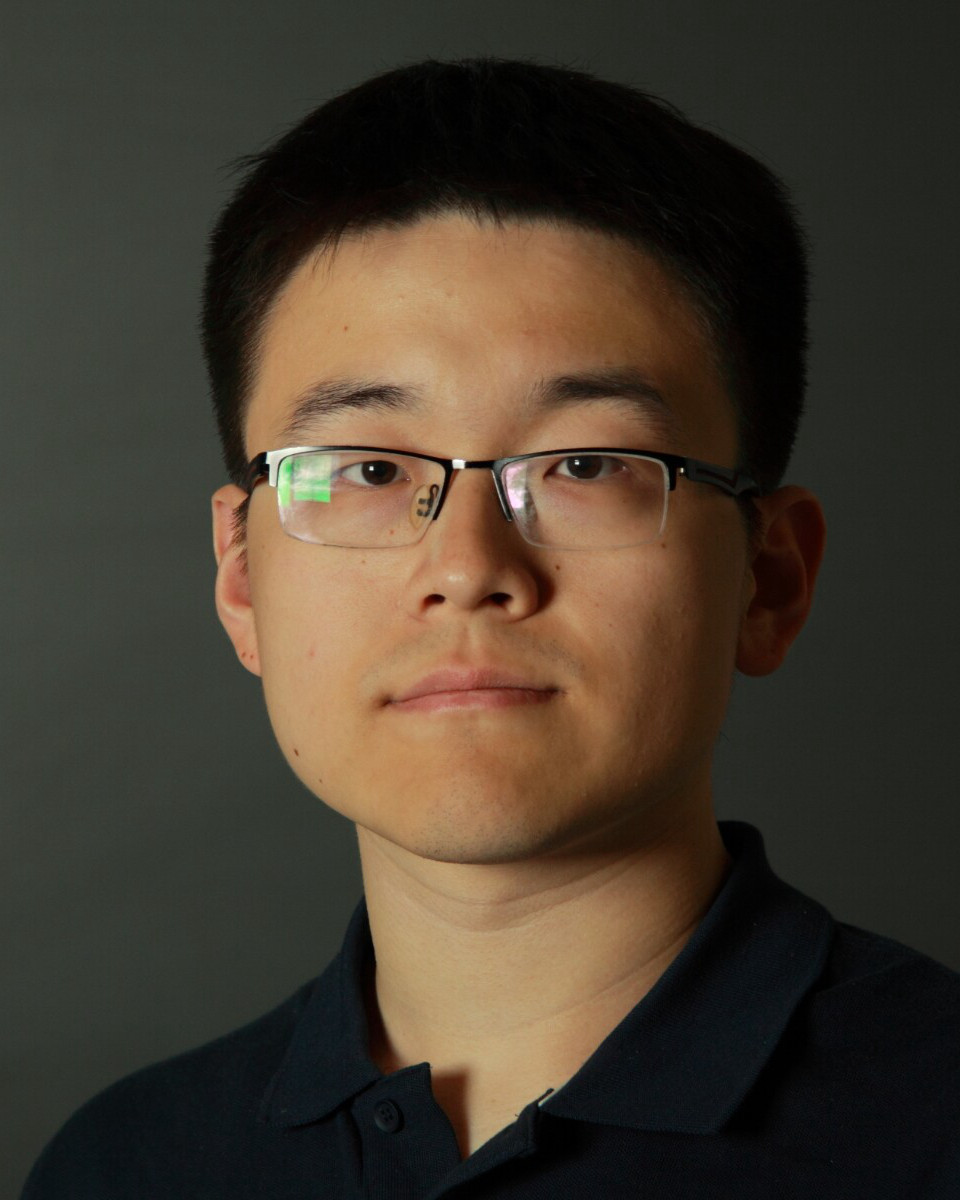}}]{Chang Gao}
received the B.Eng. degree in Electronics from University of Liverpool, Liverpool,
UK and Xi'an Jiaotong-Liverpool University, Suzhou, China, and the master’s degree in analogue and digital integrated circuit design from Imperial College London, London, UK. He is currently pursuing his Doctoral degree at the Institute of Neuroinformatics, University of Zurich and ETH Zurich, Zurich, Switzerland.
His current research interests include computer architectures for deep learning with emphasis on recurrent neural networks.
\end{IEEEbiography}
\vskip 0pt plus -1fil
\begin{IEEEbiography}[{\includegraphics[width=1in,height=1.25in,clip,keepaspectratio]{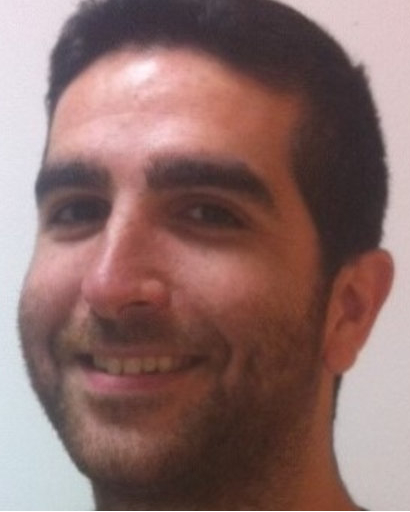}}]{Antonio Rios-Navarro}
received the B.S. degree
in computer science engineering, the M.S. degree
in computer engineering, and the Ph.D. degree in
neuromorphic engineering from the University of
Seville, Seville, Spain, in 2010, 2011, and 2017,
respectively.
He currently holds the post-doctoral position at
the Computer Architecture and Technology Department, University of Seville. His current research
interests include neuromorphic systems, real-time
spikes signal processing, field-programmable gate
array design, and deep learning.
\end{IEEEbiography}
\vskip 0pt plus -1fil
\begin{IEEEbiography}[{\includegraphics[width=1in,height=1.25in,clip,keepaspectratio]{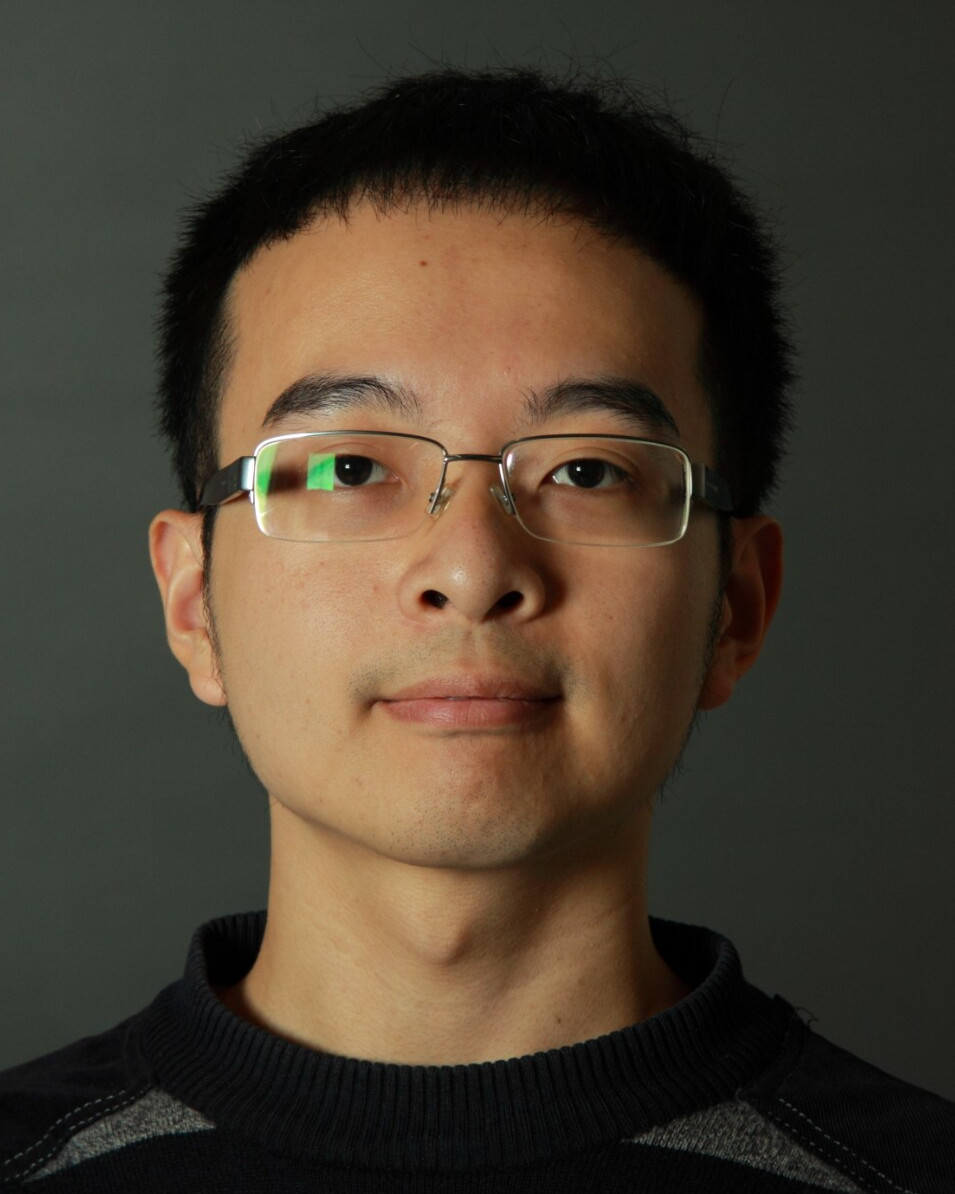}}]{Xi Chen}
received the B.Eng. degree in electrical engineering and automation from Tianjin University, Tianjin, China and in electronic and computer systems from the University of Kent, Canterbury, UK, and the M.Sc. degree in analogue and digital integrated circuit design from Imperial College London, UK. He is currently pursuing his Doctoral degree at the Institute of Neuroinformatics, University of Zurich and ETH, Zurich, Switzerland. His current research interests include hardware acceleration for deep learning and hardware architectures for on-chip training of artificial neural networks.
\end{IEEEbiography}
\vskip 0pt plus -1fil
\begin{IEEEbiography}[{\includegraphics[width=1in,height=1.25in,clip,keepaspectratio]{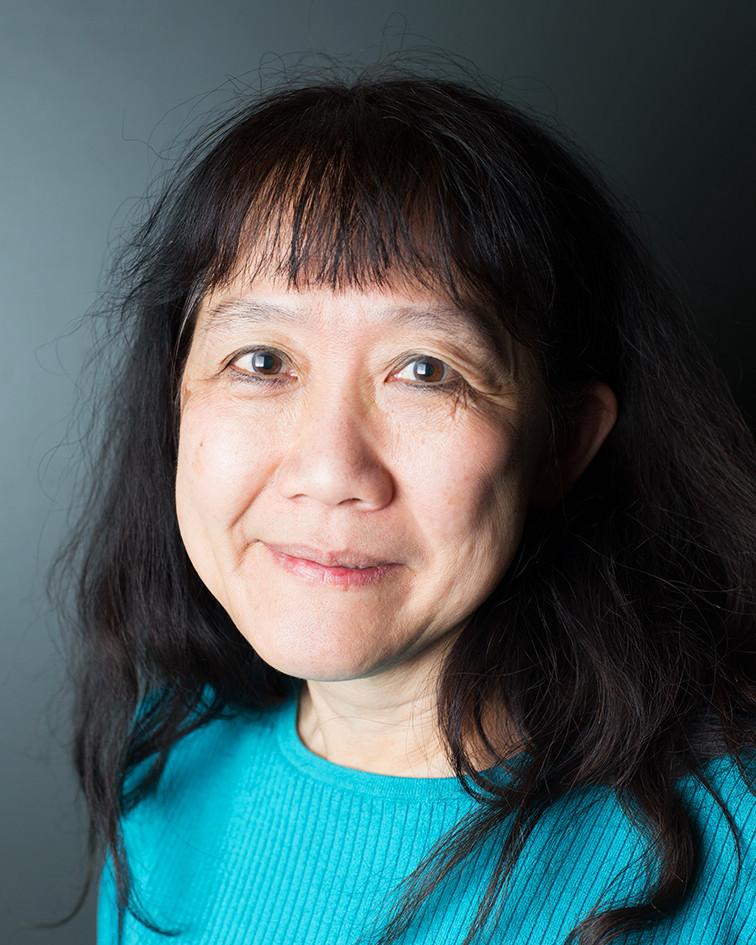}}]{Shih-Chii Liu}
(M’02–SM’07) received the bachelor’s degree in electrical engineering from the
Massachusetts Institute of Technology, Cambridge,
MA, USA, and the Ph.D. degree in the computation
and neural systems program from the California
Institute of Technology, Pasadena, CA, USA,
in 1997.
She worked in Silicon Valley before returning for her PhD.
She is currently a Professor at the University of Zurich, Zurich Switzerland. Her group focuses on audio sensors particular the spiking cochlea and bio-inspired deep neural network algorithms and hardware.
\end{IEEEbiography}
\vskip 0pt plus -1fil
\begin{IEEEbiography}[{\includegraphics[width=1in,height=1.25in,clip,keepaspectratio]{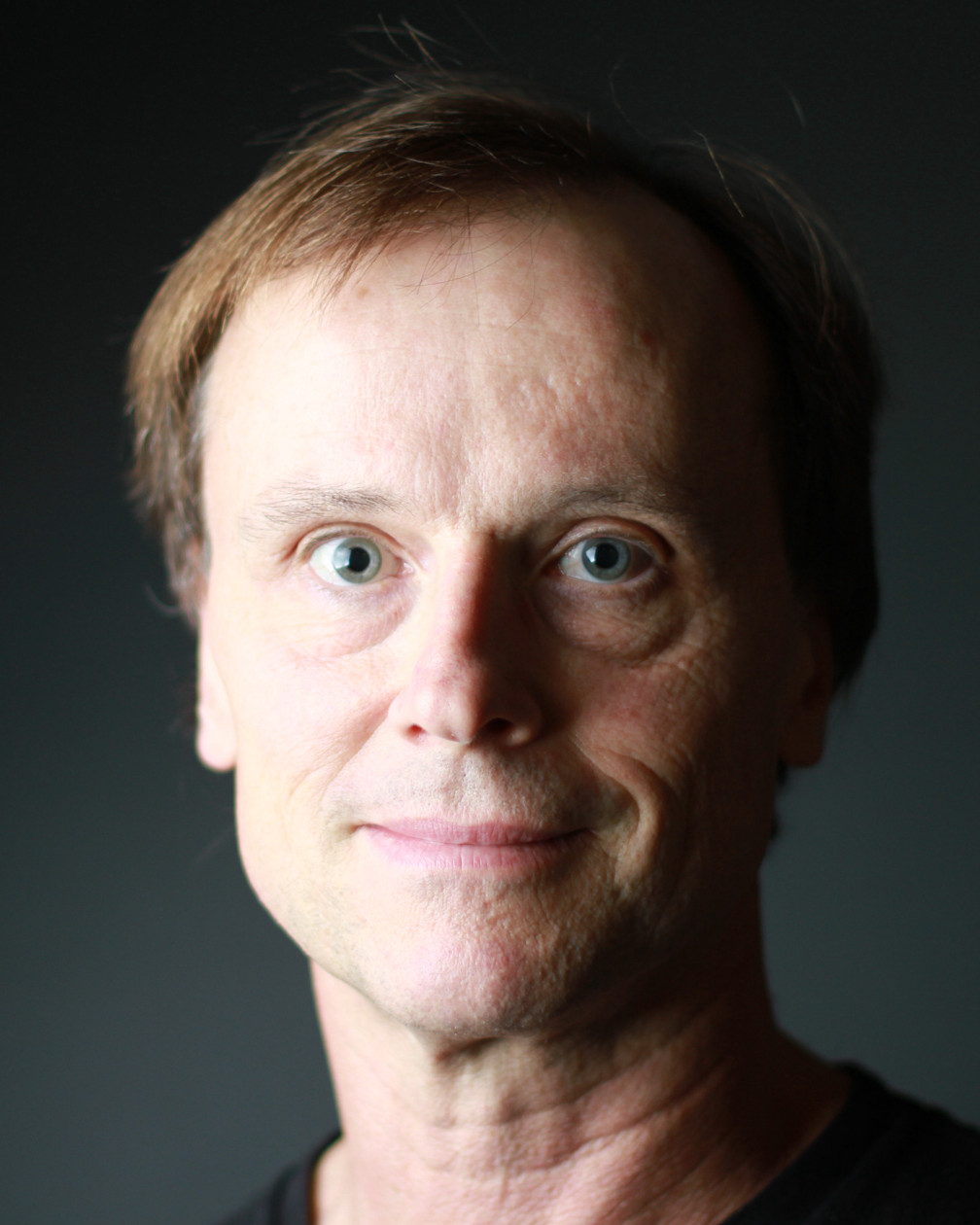}}]{Tobi Delbruck}
(M’99–SM’06–F’13) received the
B.Sc. degree in physics from the University of
California at San Diego, San Diego, CA, USA,
in 1986, and the Ph.D. degree from the California
Institute of Technology, Pasadena, CA, USA,
in 1993.
Since 1998, he has been with the Institute of
Neuroinformatics, University of Zurich and ETH
Zurich, Zürich, Switzerland, where he is currently a
Professor of physics and electrical engineering. His
group focuses on neuromorphic
sensory processing and efficient deep learning.
\end{IEEEbiography}


%\vfill


\end{document}